\documentclass[12pt]{iopart}

\usepackage{graphicx}
\usepackage{adjustbox}

\usepackage{color}
\usepackage[usenames,dvipsnames,svgnames,table]{xcolor}

\newcommand*{\paddyspeaks}[1]{\textcolor{black}{#1}}

\begin{document}

\title[The race to the bottom: approaching the ideal glass?]{The race to the bottom: approaching the ideal glass?}

\author{C. Patrick Royall}
\address{HH Wills Physics Laboratory, Tyndall Avenue, Bristol, BS8 1TL, UK.}
\address{School of Chemistry, University of Bristol, Cantock Close, Bristol, BS8 1TS, UK.}
\address{Centre for Nanoscience and Quantum Information, Tyndall Avenue, Bristol, BS8 1FD, UK.}

\author{Francesco Turci}
\address{HH Wills Physics Laboratory, Tyndall Avenue, Bristol, BS8 1TL, UK.}
\address{Centre for Nanoscience and Quantum Information, Tyndall Avenue, Bristol, BS8 1FD, UK.}

\author{Soichi Tatsumi}
\address{Kyoto Institute of Technology, Hashiue-cho, Matsugasaki, Kyoto, 606-8585, Japan.}

\author{John Russo}
\address{School of Mathematics, University Walk, Bristol, BS8 1TW, UK.}

\author{Joshua Robinson}
\address{HH Wills Physics Laboratory, Tyndall Avenue, Bristol, BS8 1TL, UK.}


\vspace{10pt}

\begin{abstract}
Key to resolving the scientific challenge of the glass transition is to understand the origin of the massive increase in viscosity of liquids cooled below their melting temperature (avoiding crystallisation).  A number of competing and often mutually exclusive theoretical approaches have been advanced to describe this phenomenon. Some posit a \emph{bona fide} thermodynamic phase to an ``ideal glass'', an amorphous state with exceptionally low entropy. Other approaches are built around the concept of the glass transition as a primarily dynamic phenomenon. These fundamentally different interpretations give equally good descriptions of the data available, so it is hard to determine which -- if any -- is correct. Recently however this situation has begun to change. A consensus has emerged that one powerful means to resolve this longstanding question is to approach the putative thermodynamic transition sufficiently closely, and a number of techniques have emerged to meet this challenge. Here we review the results of some of these new techniques and discuss the implications for the existence -- or otherwise -- of the thermodynamic transition to an ideal glass.
\end{abstract}

\maketitle

\section{Introduction}
\label{sectionIntroduction}

Glass is among the most enduring materials of everyday life, and continues to enjoy new applications, from the covers of our mobile phones \cite{mauro2014}, core of electrical transformers \cite{ediger2017} to advanced non-volatile computer memory \cite{lencer2008,wuttig2007}. Yet, the dramatic dynamic slowdown encountered upon cooling a glassforming liquid, with its 14 orders of magnitude increase in viscosity/structural relaxation time -- that is termed \textit{glass transition} -- continues to elude our understanding. Among the key questions is whether or not this massive increase in viscosity is accompanied by some kind of underlying \textit{thermodynamic} phase transition  \cite{berthier2011,cavagna2009}. Some theories assume such a transition \cite{adam1965,goldstein1969,lubchenko2007} while others are based on an \emph{avoided} transition \cite{tarjus2005} or assume that the glass transition is a predominantly \textit{dynamical} phenomenon \cite{chandler2010}. A number of excellent reviews \cite{berthier2011,cavagna2009,debenedetti,dyre2006,stillinger2013} and shorter perspectives \cite{berthier2016pt,biroli2013,debenedetti2001,ediger2017,ediger2012} of the glass transition from a theoretical viewpoint have appeared recently. More specific reviews focus on certain theoretical aspects such as the energy landscape \cite{goldstein1969,sciortino2005,heuer2008}, dynamic heterogeniety  \cite{berthier,ediger2000}, Mode-Coupling Theory  \cite{charbonneau2005,goetze}, random first-order theory (RFOT) \cite{lubchenko2007}, replica theory \cite{charbonneau2017,parisi2010}, dynamic facilitation \cite{chandler2010}, topological constraints \cite{gupta2009} and soft glassy rheology \cite{sollich2006} or materials such as polymers \cite{mckenna2017} or metallic glassformers \cite{cheng2011}.

If such an underlying phase transition exists, it may be related to the emergence of an ``ideal glass'', which would reside at the bottom of the free-energy landscape (excluding crystal configurations which represent the true minimum of the landscape). By ``ideal glass'', we refer to the counterintuitive idea that at sufficiently low (but finite) temperature the entropy of the supercooled liquid falls below the entropy of the crystal \cite{berthier2011}. This putative transition temperature is typically determined by extrapolation to low temperatures.

 At best, the ideal glass can be approached asymptotically upon cooling a liquid, due to the (presumed) divergence of the relaxation time (and hence the time required to equilibrate the material) \footnote{Except for perfectly strong liquids whose relaxation exhibits an Arrhenius dependence on temperature.}. Recently, considerable developments have been made regarding how close this (putative) state can be reached. Indeed, some experiments are now within 4\% of the temperature at which the ideal glass would be found \cite{kearns2008,ramos2011, tatsumi2012}. Here we review some of these recent developments, both experimental and computational, in techniques to approach the bottom of the energy landscape.

\emph{Scope of this work --- } The intense activity in this field of investigating very deeply supercooled states in a vast range of materials (from metals to molecules to colloids, not to mention many ingenious simulation approaches) means that any kind of comprehensive review is a major challenge. Where possible, we have referenced relevant review papers, but in any case we humbly ask for patience on the part of readers regarding those papers we have missed, or where our opinion seems at odds with theirs. This review is organised into four sections. In the following, section \ref{sectionBackground}, we briefly summarise the state--of--the--art of the field and provide contextual background. We then review recent developments in experiments in section \ref{sectionExperiments} before turning to computer simulations in section \ref{sectionSimulation}. We summarise and conclude our discussion in section \ref{sectionWhatDoesItAllMean}.

\begin{figure}
\centering
\includegraphics[width=80mm]{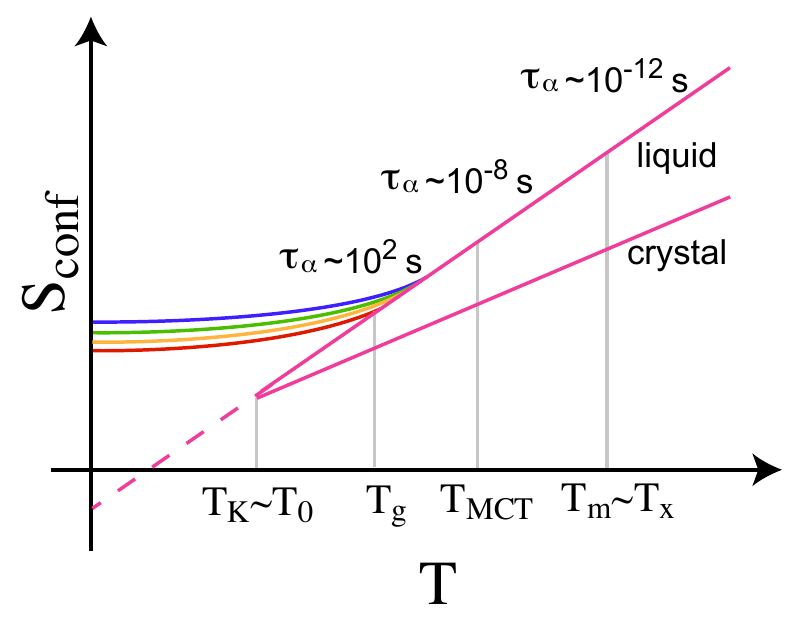}
\caption{\label{figCavagna} Roadmap to the glass transition. Configurational entropy $S_\mathrm{conf}$ as a function of temperature. Typically, the entropy of liquids falls faster than that of crystals as as function of temperature. This suggests that, at some low temperature --- the Kauzmann Temperature $T_{K}$ --- the liquid entropy would fall below that of the crystal. $T_K$ is numerically close to $T_0$, the temperature at which the structural relaxation time $\tau_\alpha$ is predicted to diverge by the Vogel-Fulcher-Tamman expression Eq. ~\ref{eqVFT}. $T_{g}$ is the operational glass transition temperature where the structural relaxation time reaches 100 s. $T_{\mathrm{mct}}$ is the mode-coupling transition, $T_m$ is the melting point and $T_{x}$ denotes a crossover temperature below which relaxation occurs through cooperative motion, \emph{i.e.} the energy landscape becomes significant.}
\end{figure}

\section{Background : paradigm for the ideal glass}
\label{sectionBackground}

\textit{What is the glass transition? ---} 
The nature of the glass transition, if any, is not understood. On timescales accessible to experiment and computer simulation, the glass transition takes the form of a \emph{continuous} increase in viscosity, or structural relaxation time. In normal liquids, the relaxation occurs on the picosecond timescale. When the relaxation time reaches 100s, a supercooled liquid is termed a \textit{glass}. This is a purely operational definition (it is easier to time 100 s than a divergent timescale!) which can obscure the fact that at the temperature corresponding to a relaxation time of 100 s, $T_{g}$, \emph{precisely nothing} happens in a thermodynamic sense, so $T_g$ does not correspond to any kind of true glass transition. Put another way, $T_{g}$ could as well correspond to a relaxation time of 10 s or 1000 s. Thus whether we have a supercooled liquid or a glass depends, not necessarily on the material or state point, but upon how long we wait.

That the structural relaxation time grows to unmanageable timescales has significant consequences. The relaxation time is an indicative measure of how long it takes a liquid to reach equilibrium. At least 10-100 relaxation times are required before one can say that a liquid is ``in equilibrium''. As the relaxation time reaches 100s, a material will appear solid. It will no longer flow on reasonable timescales. This gradual \emph{emergence} of solidity that characterises the glass transition can obscure a fundamental question. Because it is very hard to equilibrate the material below $T_g$, it is very hard to answer questions about whether there is a true thermodynamic transition to an ideal glass at some lower temperature $T_K<T_{g}$. It is this equilibration issue that we seek to address here.

\begin{figure}
\centering
\includegraphics[width=80mm]{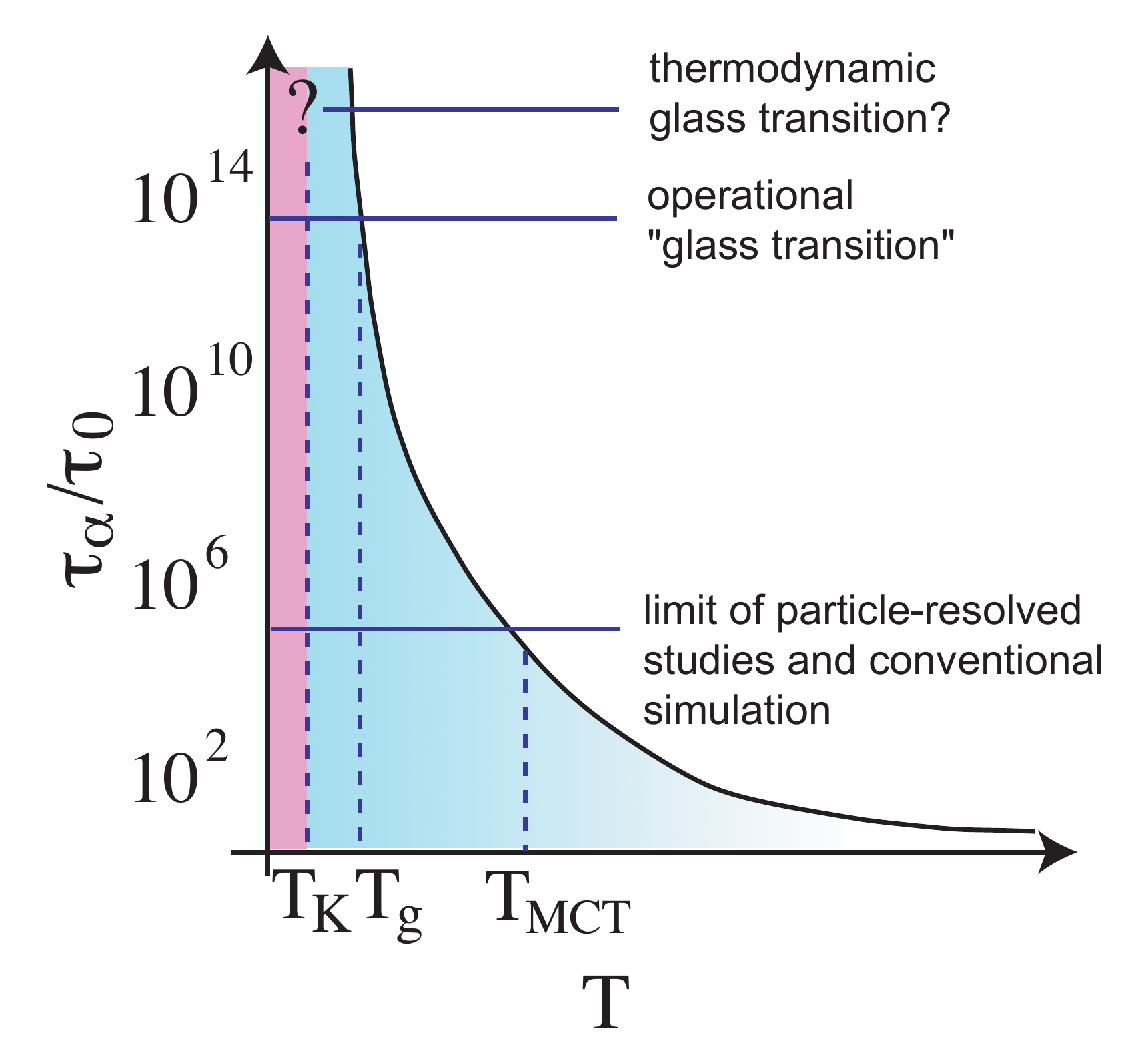}
\caption{\label{figGlassChallenge} 
The challenge of the glass transition. The operational glass transition, $T_g$ is the limit of most experiments, some 14 decades in relaxation time slower than the normal liquid. This still doesn't reach the putative ideal glass transition at the Kauzmann temperature $T_K$. Conventional computer simulations and particle-resolved colloid experiments are hampered even further, as they are limited to only around five decades.}
\end{figure}

\textit{The ideal glass --- }
It is suggested in some theoretical approaches \cite{berthier2011,adam1965,lubchenko2007,parisi2010} that there is a thermodynamic transition at or around the Kauzmann temperature $T_K$ where upon extrapolation, the entropy of the supercooled liquid falls below that of the crystal \cite{kauzmann1948}. We emphasise that it is quite possible that this paradox, of an amorphous state having a lower (or equal) entropy than the crystal, may be avoided, for example by crystallisation \cite{kauzmann1948,tanaka2003} or by other effects, such as thermal defects \cite{stillinger2001}. Indeed, arguments have been made against the existence of the \paddyspeaks{so-called} Kauzmann paradox \cite{chandler2010,hoffmannw2012,aliotta2017}. \paddyspeaks{In particular, it is important to re-iterate the point made in the preceding section: any such ideal glass is not expected to represent the true minimum of the energy landscape, as that would correspond to the crystal \cite{hoffmannw2012}. Other possibilities that the ideal glass might be avoided include thermal defects \cite{stillinger2001}, or that it is not even expected in theories such as dynamic facilitation \cite{chandler2010}. 
However, a number of theoretical approaches, Adam-Gibbs and RFOT amongst them \cite{adam1965,lubchenko2007} invoke a drop in the configurational entropy in the amorphous state as a mechanisms for vitrification. In this way of thinking, the system is presumed not to undergo crystallisation and thus the true bottom of the energy landscape, corresponding to the crystal, is neglected. Rather than comparing the entropy with respect to the crystal, as Kauzmann did in his original 1948 paper \cite{kauzmann1948}, in much of the work covered in this review, we are interested in the entropy in the amorphous state only.}
For our present purposes, we note that the supposed existence of this 
Kauzmann point, existent or otherwise, nevertheless serves as a useful landmark when considering the behaviour of glassforming systems \cite{ediger2017,berthier2011,cavagna2009,goldstein1969,lubchenko2007,debenedetti,dyre2006,stillinger2013,berthier2016pt,biroli2013,debenedetti2001,ediger2012,sciortino2005,heuer2008,charbonneau2017,parisi2010}, and we therefore use the Kauzmann point to guide our approach to deeply supercooled states low in the energy landscape.

 Often, the approach to the bottom of the energy landscape is thought of in terms of the \textit{configurational} entropy of the supercooled liquid, which is expected to become very small \cite{adam1965,lubchenko2007}. For our purposes, the \emph{configurational entropy} $S_\mathrm{conf}$ is defined as $S_\mathrm{conf}=S-S_\mathrm{vib}$, where $S$ is the total entropy of the system, and $S_\mathrm{vib}$ is the vibrational entropy due to thermal excitations around inherent states, \emph{i.e.} the energy minima that are reached following a steepest descent quench. Note that the vibrational entropy $S_\mathrm{vib}$ also implicitly includes contributions from the entropy of mixing in the case of multicomponent systems \cite{sciortino1999}. $T_K$ is defined as the temperature at which $S_\mathrm{conf}$ becomes sub-extensive, meaning that it vanishes in the thermodynamic limit. The Kauzmann paradox observes that the liquid entropy falls faster than that of the crystal and at a finite temperature $T_K<T_{g}$, the two should cross~\cite{kauzmann1948}.

Here we focus on the configurational entropy, $S_\mathrm{conf}$ (Fig. \ref{figCavagna}). While we follow the ideas of Ref.  \cite{sciortino1999}, we emphasise that this splitting of the entropy into vibrational and configurational parts is valid provided there is a sufficient decoupling between vibrational timescales ($\beta$-relaxation) and timescales of full relaxation ($\alpha$-relaxation). This implies a certain degree of supercooling \cite{sciortino1999}.  However, this may be operationally challenging \cite{angell2002}, and recent computer simulations have explored different ways to determine the configurational entropy (see Section \ref{sectionNonLocal}) ~\cite{berthier2017pnas} depending on the precise definition of the vibrational contribution. Given that the configurational entropy of the crystal is small, this suggests that the ideal glass should have an exceptionally low configurational entropy while nevertheless remaining amorphous (Fig. \ref{figCavagna}). Alternatively, there may be no transition until absolute zero temperature~\cite{cavagna2009,berthier2011,chandler2010}. The focus of this work is to review recent progress towards addressing this conundrum. 

Another significant point is the mode-coupling temperature at $T_\mathrm{mct}$ where, according to this particular mean-field theory, a divergence of the relaxation time is predicted to occur~\cite{charbonneau2005}. In its usual implementation, mode-coupling theory does not include the effect of activated processes so its associated dynamical divergence is never truly observed \footnote{Extensions to MCT which have the potential to include collective relaxation include inhomogeneous MCT~\cite{biroli2006}, Generalised MCT ~\cite{janssen2015} and hopping, which enables a link to be made between MCT and random first-order transition theory , thus connecting MCT with theories which make a direct link to the energy landscape ~\cite{bhattacharyya2005,bhattacharyya2008}.}. But echoes of this mean field transition can be observed in experiment, and it also marks the emergence of activated relaxation.

In this discussion, we have identified \emph{three} relevant temperatures : $T_g$ the experimental transition, $T_K$ the possible thermodynamic transition to the ideal glass, and $T_\mathrm{mct}$ the mode-coupling temperature. An important point is that in considering dynamical arrest, one should think of glass transition-\emph{s}, or that at least we must specify what we mean by ``glass transition''. Here we focus on the approach to $T_K$. We now briefly review some key concepts concerning the behaviour of systems as they move low in their free energy landscape. We note that recently, a further transition in amorphous systems has been  identified. This \emph{Gardner transition} occurs between two glass states, but is not pertinent to our discussion of the quest for the ideal glass \cite{charbonneau2017,charbonneau2014}.

\subsection{Fragility and the Angell Plot}
\label{sectionFragility}

\begin{figure}[!htb]
\centering 
\includegraphics[width=80mm]{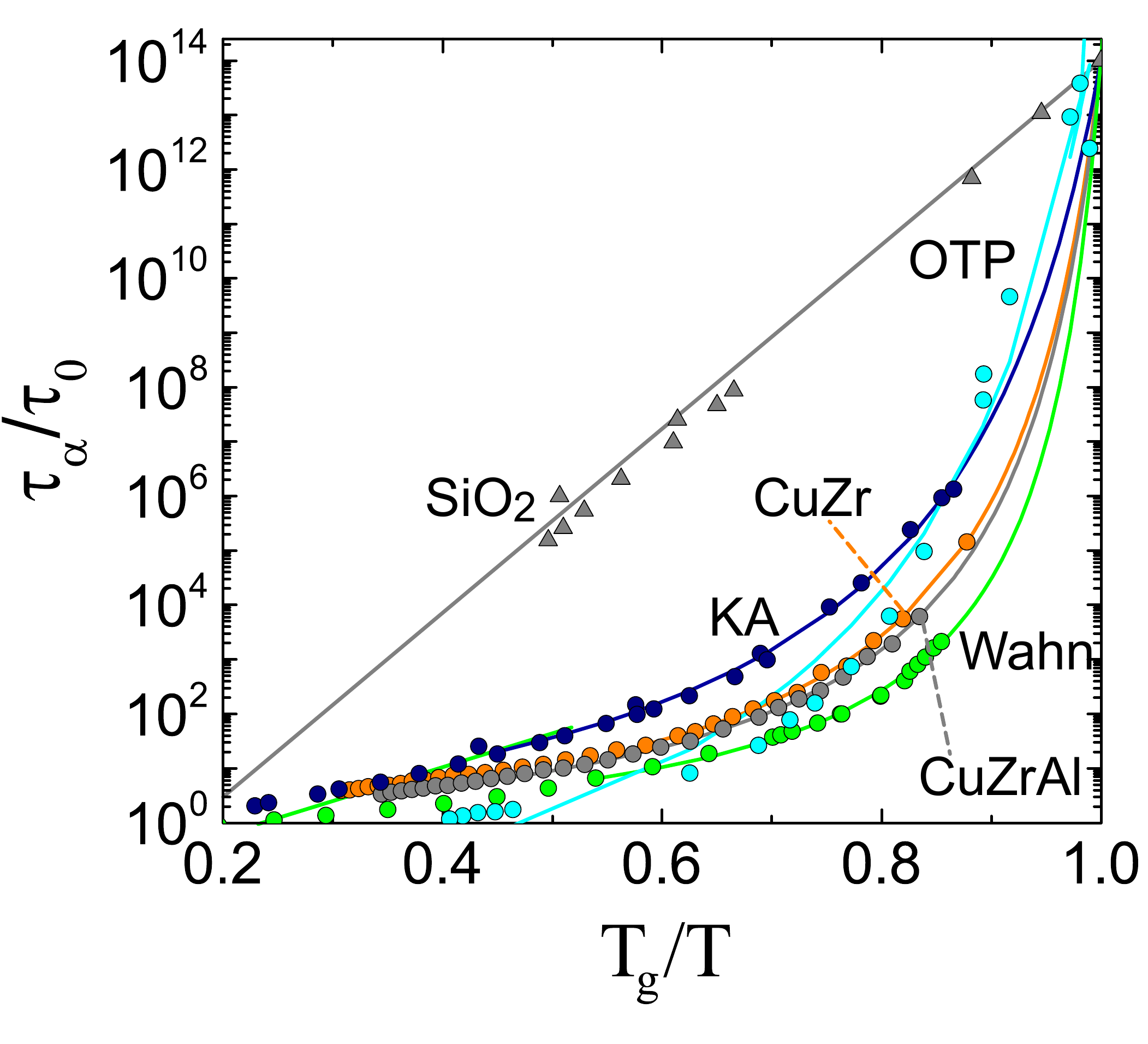} 
\caption{The Angell plot : Arrhenius representation of liquid viscosities, with inverse temperature scaled by $T_{{\rm g}}$. \textit{Strong liquids} exhibit Arrhenius behaviour characterised by an approximately straight line, indicative of a temperature-independent activation energy. \textit{Fragile liquids}, on the contrary, reveal super-Arrhenius behaviour, where activation energy grows as temperature decreases \cite{angell1988}. Data for SiO$_2$ and orthoterphenyl (OTP) are quoted from Angell \cite{angell1995} and Berthier and Witten \cite{berthier2009witten}. The other data concern model systems: KA denotes Kob-Andersen and Wahn denotes Wahnstr\"{o}m binary Lennard-Jones systems respectively while HS denotes hard spheres where the control parameter is the compressibility factor $Z$  \cite{berthier2009,royall2017}. CuZr and CuZrAl denote numerical calculations of metallic glassformers by ourselves based on an embedded atom model for Cu$_{65}$Zr$_{27.5}$Al$_{7.5}$~\cite{robinson2017}. The latter matches the composition of the experiments by Yu {\it et al}.\cite{yu2013}. Here, $\tau_0$ is scaled to enable data collapse at $T_g/T = \phi/\phi_g$ \cite{royall2015}. }
\label{figAngellEternal} 
\end{figure}

The Angell plot~\cite{angell1988} offers a convenient representation of the dynamical slowing down underlying the glass transition, and highlights some of the most important differences between glassforming materials. In this plot, the logarithm of the structural relaxation time $\tau_\alpha$ (or the viscosity) is plotted against $T_g/T$. In this representation, an Arrhenius-like behaviour,

\begin{equation}
\tau_{\alpha}\propto \exp\left( \frac{E_{\mathrm{a}}}{k_{\mathrm{B}}T} \right),
\label{eqArr}
\end{equation}

\noindent where $E_{\mathrm{a}}$ is viewed as an activation energy, should be represented by a straight line. In Fig.~\ref{figAngellEternal}, we plot the logarithm of the viscosity, which (neglecting Stokes-Einstein breakdown \cite{berthier2011}) is proportional to $\tau_{\alpha}$, showing that for some materials, notably silica, it is indeed close to a straight line. However, for many supercooled liquids a \emph{fragile} or super-Arrhenius behavior is found. The increase in relaxation times is often well described by the semi-empirical Vogel-Fulcher-Tamman (VFT) expression,

\begin{equation}
\tau_{\alpha}=\tau_{0}\exp\left[ \frac{DT_{0}}{T-T_{0}} \right]
\label{eqVFT}
\end{equation}

\noindent where $T_{0}~(\simeq T_{{\rm K}}$) is rather lower than the experimental glass transition temperature $T_{{\rm g}}$ and $D$ is a measure of the \textit{fragility} -- the degree to which the relaxation time increases as $T_g$ is approached. Smaller values of $D$ correspond to more fragile materials. Furthermore, Eq.~\ref{eqVFT} can be rationalised by both Adam-Gibbs \cite{adam1965} and random-first order transition \cite{lubchenko2007} theories. Remarkably, $T_0 \approx T_K$ for a large number of glassformers, fuelling the idea of a thermodynamic phase transition at $T_K$ \footnote{See Cavagna  \cite{cavagna2009} for an interpretation of $T_0 \approx T_K$}. For our purposes, let us emphasise that for a large number of materials, $T_0$ and $T_K$ vary from one another by 10\%. While it is worth noting (see earlier, Section \ref{sectionBackground}) that the accurate determination of $T_K$ (and $T_0$) is not without its challenges, more fundamentally, one expects a significant connection with fragility. It is suggested that more fragile materials exhibit more change in structure, and thus their configurational entropy should also change more~\cite{martinez2001} (see also section \ref{sectionDynamicStaticLengthscales}). Conversely, less fragile (\emph{i.e. stronger}) materials exhibit a smaller change in structure, and indeed are rather poorly fitted by the VFT expression Eq. \ref{eqVFT}, so it comes as no surprise that strong materials such as silica exhibits differences in $T_K$ and $T_0$ of 15\% \cite{tanaka2003prl}. Some polymers, notably polycarbonate and polyamide exhibit more substantial deviations (the latter has a $T_0$ over 30\% above its $T_K$) \cite{cangiolosi2005}. Nevertheless, very many glassformers do have $T_0$ and $T_K$ close together.

As successful as Eq.~\ref{eqVFT} has been (at least for fragile glassformers), it is not the only option. In fact, as Tab.\ref{tableHowLow} indicates, the value of $T_K$ 
deviates from $T_{0}$. Dyre and co-workers showed that VFT is not the best fit to experimental data \cite{hecksler2008}, and other forms, for example one also related to the Adam-Gibbs theory \cite{mauro2009}, provide better agreement with experimental data. Furthermore, the Elmatad-Garrahan-Chandler form \cite{elmatad2009}, based on dynamic facilitation theory \cite{chandler2010} fits the experimental data just as well, and assumes no dynamic divergence at finite temperature. Other examples in which no dynamic divergence is encountered include expressions such as that developed by Mauro \emph{et al.} \cite{mauro2009} which relate configurational entropy the topological degrees of freedom per atom, elastic treatments of excitations which enable relaxation \cite{mirigian2013}, geometric frustration \cite{tarjus2005} and related models which treat glassformers with a population dynamics model of locally favoured structures \cite{pinney2015}. Nevertheless, here we shall use the VFT form and presume that the divergence at $T_0$ is representative of $T_K$ \cite{cavagna2009}, for ease of comparison with earlier work \cite{swallen2007}.

\subsection{Dynamic and static lengthscales}
\label{sectionDynamicStaticLengthscales}

\begin{figure}[!htb]
\centering \includegraphics[width=85mm]{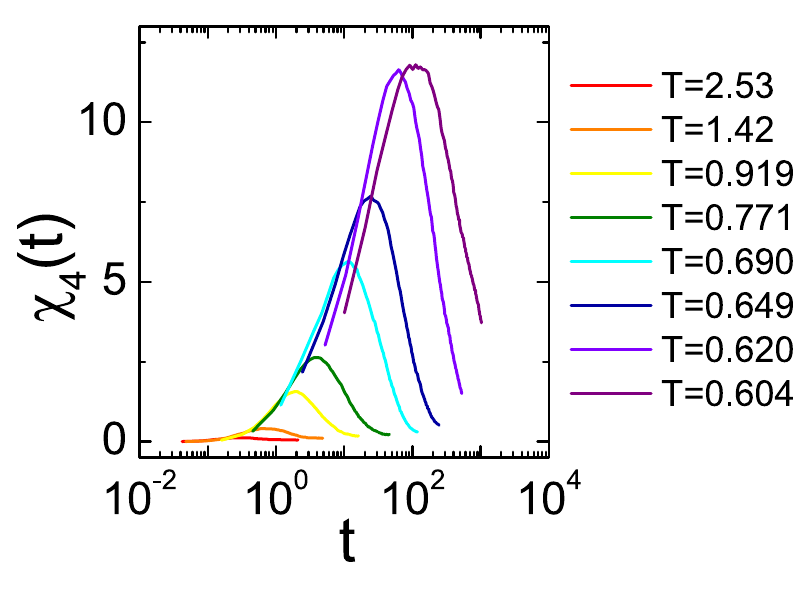} 
\caption{The dynamic susceptibility $\chi_{4}(t)$ in the Wahnstr\"{o}m binary Lennard-Jones glassformer. Colours correspond to different temperatures
\cite{malins2013jcp}. }
\label{figChi4} 
\end{figure}

\textit{Dynamic Heterogeneity --- }
Approaching the glass transition, \textit{dynamic heterogeneity} becomes a significant feature: Particles move in an increasingly cooperative manner creating dynamically correlated mesoscopic domains \cite{perera1996,kob1997}. A \textit{dynamic length scale} can be associated with the increasing \index{dynamic heterogeneity} dynamic heterogeneity -- a measure which characterizes the size of growing cooperative motion. This lengthscale can be quantified by using so-called four-point correlation functions.
The fluctuations associated with dynamic heterogeneity are then characterised by the \emph{dynamic susceptibility},
\begin{equation}
\chi_{4}(t)=\frac{1}{N \rho A_{{\rm B}}T}\left[\langle W^{2}(t)\rangle-\langle W(t)\rangle^{2}\right],
\label{eqChi4}
\end{equation}
\noindent which is obtained by integrating a four-point time-dependent density correlation function over volume \cite{lacevic2003}. For a given temperature $T$ (or packing fraction $\phi$), the fluctuations (\emph{i.e.}, the susceptibility) attain a maximum at certain time
$t=\tau_{h}$  $(\approx \tau_\alpha)$, and then die away, as illustrated in Fig.~\ref{figChi4}. The time where $\chi_{4}(t)$ is maximised $\tau_{h}$ has a value similar to $\tau_{\alpha}$.

The exact nature of the increase of the dynamic length scale remains unclear, and depends on the system under consideration.  In simulations of the Wahnstr\"{o}m binary Lennard-Jones glassformer, this increase of a dynamic lengthscale was likened to divergence of the static density-density correlation length observed in liquid-gas critical phenomena \cite{lacevic2003}. However it has also been claimed that the increase of the dynamic lengthscale related to $\chi_4$, $\xi_4$ supports the dynamic facilitation interpretation (divergence of the dynamics only at $T=0$) \cite{whitelam2004}, while other work points to a divergence at the VFT temperature $T_0$ \cite{tanaka2010}. Recent work involving deep supercooling (by the standards of computer simulation) and careful analysis indicates a change in scaling of $\xi_4$ around the mode-coupling crossover \cite{flenner2013}. Such a dynamic lengthscale cannot be measured directly in experiments on molecules. However indirect measurements have been made in a variety of ways, all arriving at similar conclusions: the size of the cooperatively rearranging regions (CRRs) is only around 10-100 molecules at $T_g$, where the dynamics are ten decades slower than around the mode coupling temperature, close to the limit of supercooling that can be reached via simulations \cite{ediger2000,tatsumi2012,donth1982,cicerone1995,berthier2005,tracht1998,ashtekar2012,dalleferrier2007,yamamuro1998}. We are thus left with the unsettling situation that \emph{simulations and experiments obtain similar values for the dynamic length when the dynamics differ by ten decades.} These studies underline the difficulties of obtaining sufficiently supercooled samples where the lengthscale changes enough to enable a clear discrimination to be made.

\textit{The rationale for a static lengthscale --- } Building on ideas introduced by Bouchard and Biroli, \cite{bouchaud2004}, Montanari and Semerjian considered a cavity made of immobile particles outside a certain radius, and where re-arrangements are possible inside~\cite{montanari2006}. Now, above a certain cavity size, particle re-arrangements are just possible, and thus $\tau_\alpha$ is finite. In other words, there is a lengthscale, above which there is more than one way to arrange the particles in the cavity. Under these conditions, the system \emph{must} be able to relax and a maximum relaxation time $\tau_\alpha^{\mathrm{max}}$, which corresponds to the transition between these different configurations. This cavity has a fixed lengthscale, and, under these assumptions, $\tau_\alpha^{\mathrm{max}}$ should be Arrhenius, as we have that there are multiple configurations inside the cavity and there is some activation barrier --- whose height fixes $\tau_\alpha^\mathrm{max}$, \emph{i.e.} there is some fixed maximum activation energy $E^\mathrm{max}_a$ in Eq. \ref{eqArr}.  Of course, especially in the liquid at temperature $T$ above the melting temperature $T_m$, the system relaxes much faster than $\tau_\alpha^{\mathrm{max}}$. However, if $\tau_\alpha$ follows a super-Arrhenius behaviour, it must, eventually, cross $\tau_\alpha^{\mathrm{max}}(T)$. This would mean that the assumption of a cavity with the number of available configurations constant with respect to temperature is flawed. \emph{In other words, under super-Arrhenius dynamics some structural change must occur. i.e. the system moves lower in its energy landscape.}

In addition to the arguments above, theories which imagine a thermodynamic origin to the glass transition \cite{adam1965,lubchenko2007,tarjus2005,parisi2010} predict that static lengthscales should increase concurrently with dynamic lengthscales. In the majority of investigations (limited to the weak supercooling regime to $T_\mathrm{mct}$ accessible to computer simulations), the dynamic ($\xi_4$) and static lengths appear decoupled \cite{malins2013jcp,cammarota2012epl,cammarota2012,cammarota2013,hocky2012,dunleavy2012,malins2013fara,charbonneau2012,charbonneau2013jcp,charbonneau2013pre,hocky2014,royall2015,dunleavy2015,charbonneau2016}, with the possible exception of weakly polydisperse hard-spheres~\cite{tanaka2010,kawasaki2010jpcm,leocmach2012,leocmach2013} and and certain innovative measures of local solidity \cite{mosayebi2010}. It is possible that other dynamic lengths might resolve this issue \cite{dunleavy2015}, but in our opinion, there is a need to obtain data at deeper supercooling than has been possible hitherto. The role of lengthscales in the glass transition has recently been reviewed by Karmakar \emph{et. al} \cite{karmakar2014} and Harrowell \cite{harrowell2011}.

\subsection{The Energy Landscape}
\label{sectionEnergyLandscape}

\begin{figure*}[!htb]
\centering \includegraphics[width=125mm]{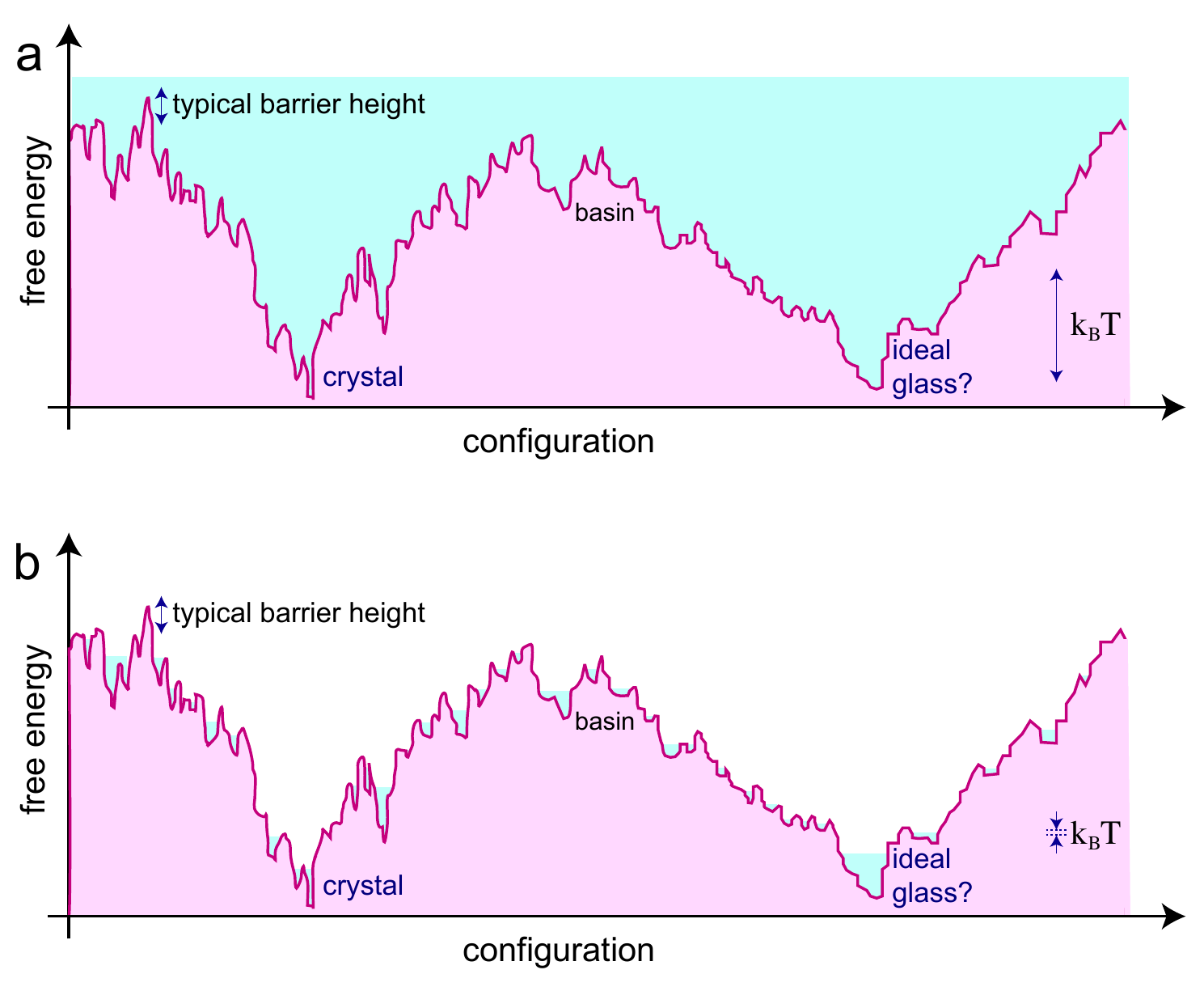} 
\caption{A schematic illustration of the potential energy landscape. The $x$-axis represents configurations of all $3N$ coordinates. 
The bottom of the energy landscape is typically the crystal, but the energy minimum for the ideal glass may be comparable.
(a) High temperature liquid, the typical barrier height is less than the thermal energy, and all configurations can be accessed as indicated in blue. (b) Low temperature glass. The barrier height between basins is now much higher than $k_BT$, and the system is no longer able to explore all configurations and has become a glass but can only access local states, as indicated by the blue shading.
\label{figEnergyLandscape} }
\end{figure*}

One very intuitive picture of the equilibrium dynamics of a deeply supercooled liquid is provided by Goldstein's energy landscape~\cite{berthier2011,goldstein1969}. Goldstein put the emphasis on the evolution of the system in phase space, \emph{i.e.} the space of all the configurational degrees of freedom. In the case of a simple liquid in three dimensions, this is the space of all 3N coordinates of the particles. The total potential energy is then $U(r_0,r_1..r_N)$ where the $r_i$ are the coordinates of the $N$ particles which comprise the system.  Each configuration is represented by a point in phase space, and the dynamics of the system can be thought of as the motion of this point over this \emph{potential energy landscape}. In other words, such a classical system is entirely described by its energy landscape \cite{cavagna2009}. The local minima of the potential energy correspond to mechanically stable configurations of the particle system. One of these is the crystal, and this will usually be the absolute minimum (Fig. \ref{figEnergyLandscape}). Beyond the crystal-related minima, there will be many local minima corresponding to particle arrangements that lack crystalline order. These are amorphous, or glassy, minima, and have a potential energy that is 
larger than the crystal one.

Below some crossover temperature ($T_x$ in Fig. \ref{figCavagna}), a supercooled liquid explores the phase space mainly through activated hops between different amorphous minima, which are separated by potential energy barriers. Note that it is in (meta) equilibrium (metastable to crystallisation), so the potential energy is constant in time, save for thermal fluctuations  (we assume there is no ageing) and the system is ergodic \cite{cavagna2009}. One should note that these hopping events involve only a few particles (the typical number of dynamically correlated particles can be estimated by $\chi_4(t)$, Eq. \ref{eqChi4}, Fig. \ref{figChi4}).

Henceforth we review recent developments in approaching the bottom of the energy landscape (neglecting crystallisation). We begin by considering experimental approaches, and move on to consider computer simulations.

\section{Approaching the bottom of the energy landscape in experiments}
\label{sectionExperiments}

\begin{figure*}[!htb]
\centering \includegraphics[width=70mm]{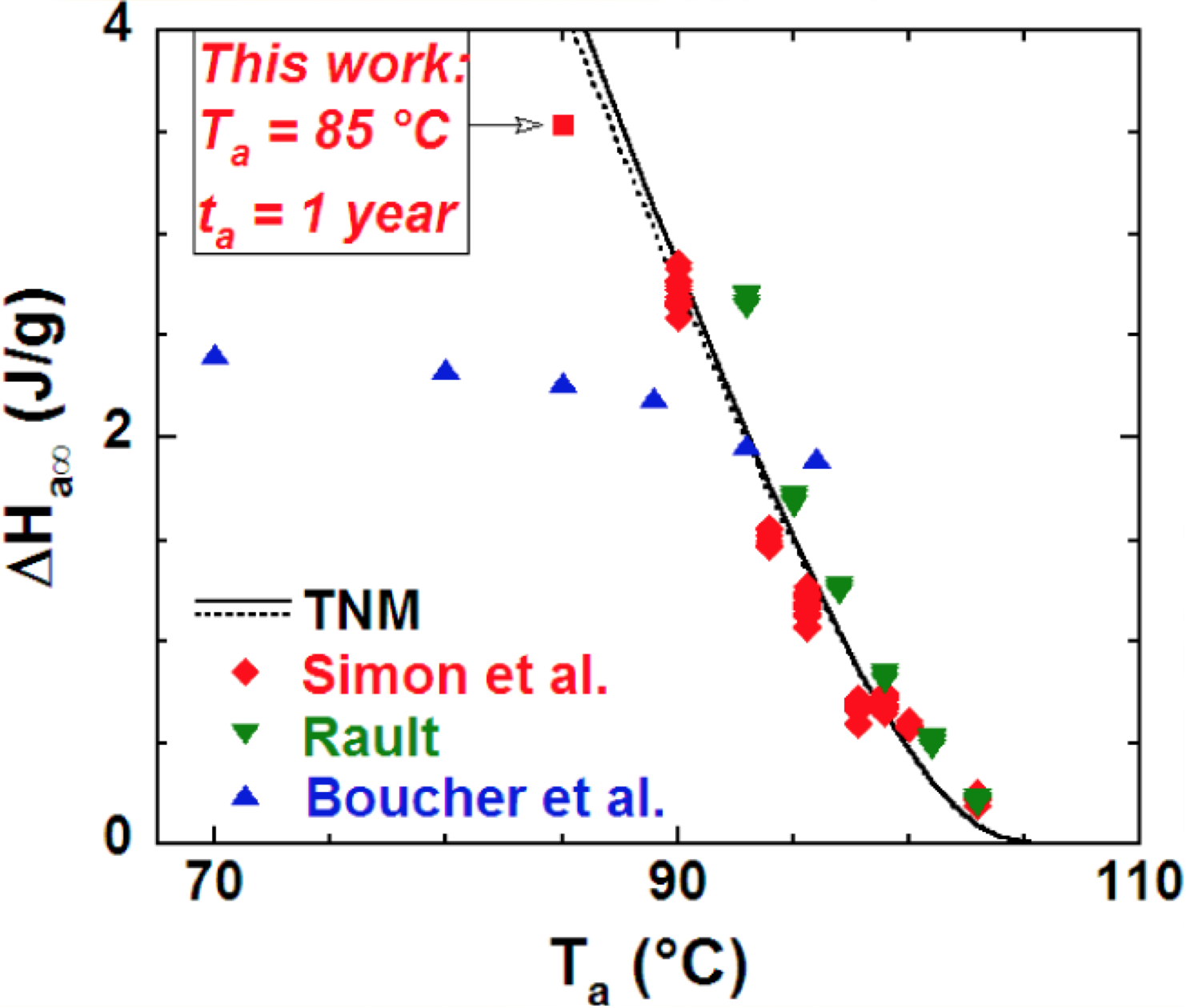} 
\caption{
Aging experiments on polystyrene glass. Enthalpy loss at equilibrium ($\Delta H_{a, {\infty}}$) as a function of temperature. Koh and Simon's work is shown as the red squares. The black solid line is a theoretical prediction from Tool--Narayanaswarmy--Moynihan  
\cite{tool1946,narayanaswarmy1971,moynihan1976} which takes into account the temperature dependence of change in heat capacity $\Delta C_p$, and the black dotted line is the prediction with fixed value of  $\Delta C_p$ at $T_g$.  The data point from this work is a lower bound. The key point of this figure is that the $\Delta C_p$ of Koh and Simon's work corresponds to that of an equilibrium supercooled liquid 4\% below $T_g$ in absolute terms. Reprinted with permission from Y. P. Koh and S. L. Simon. \emph{Macromolecules}, \textbf{46} 5815--5821 (2013). Copyright 2013 American Chemical Society.
\label{figKohSimon} }
\end{figure*}

\begin{table*}[t]
\begin{adjustbox}{width=1.\textwidth,center}
\begin{tabular}{ | c || c | c | c | c | c | c | c | c | } \hline
system	& $T_K$, $Z_K$ 			& $T_0$, $Z_0$ 				& $T_g$, $Z_g$		& $T_\mathrm{eq}$, $Z_\mathrm{eq}$	& $D$				& $\theta_0$			& $m$	 						& Technique \\ \hline\hline
EB 		& 101\cite{tatsumi2012} 		& 95.5\cite{chen2011a} 			& 116 				& 105\cite{ramos2011}				& 8.30				& 0.53 				& 97.5\cite{chen2011a}				& VD\cite{ramos2011}\\
ECH 	& 76\cite{chua2015}	 	& 67.1\cite{mandanici2008a}		& 101 				& 94\cite{ramos2015} 				& 21.5 				& 0.19 				& 56.5\cite{mandanici2008} 			& VD\cite{chua2015,ramos2015}\\
TNB 	& 270\cite{magill1967} 		& 264\cite{richert2003}			& 345				& 306\cite{swallen2007}				& 14.1				& 0.48 				& 86\cite{richert2003}				& VD\cite{swallen2007}\\
TL 		& 108\cite{tatsumi2012} 		& 98.2\cite{dos1997}			& 117 				& 108\cite{sepulveda2011}			& 7.51				& 0.48 				& 104\cite{chen2011a} 				& VD\cite{sepulveda2011}\\
IMC 		& 240\cite{shamblin1999}	& 234\cite{wojnarowska2009}		& 314 				& 286\cite{swallen2007}				& 					& 0.36 				& 82.8\cite{wojnarowska2009}		& VD\cite{swallen2007}\\
PR 		& 49.8\cite{tatsumi2012} 		&							& 56\cite{tatsumi2012}	&  								&  					&  					&  								& VD\cite{tatsumi2012}\\
CCl$_4$ 	&  						& 69\cite{chua2016} 			& 78.5\cite{chua2016}	&	 							& 4.54				& 	 				& 118\cite{chua2016}				& VD\cite{chua2016}\\
MG 		&  						& 625\cite{robinson} 			& 676\cite{yu2013}		& 640\cite{yu2013} 					& 2.85 \cite{robinson} 	& 0.70 				& 38\cite{yu2013} 					& VD\cite{yu2013}\\  \hline
PS 		&  						& 321\cite{PolymerHandbook2007}& 373 				& 361\cite{koh2013}				& 6.17				& 0.23 				& 116\cite{PolymerHandbook2007}		& $t_\mathrm{w} \sim 1$ year\cite{koh2013}\\
PCB 	& 290\cite{cangialosi2004}	& 374\cite{PolymerHandbook2007}& 424 				& 408\cite{boucher2011} 			& 3.37 				& 0.31 				& 91.6\cite{PolymerHandbook2007}	& $t_\mathrm{w} \sim 1$ year\cite{boucher2011}\\
AMB 	& 						& 338\cite{zhao2014thesis}		& 409 				& 366\cite{zhao2013}				& 7.29				& 0.61 				& 87\cite{zhao2014thesis}			& $t_\mathrm{w} \sim 2\times 10^7$ year\cite{zhao2013}\\  \hline
OTP 	& 225\cite{yamamuro2012}	& 202\cite{richert1998} 			& 246 				& 230 \cite{aliotta2017}				& $\sim 10$  \cite{richert1998}   	& 0.363 
																																		\cite{aliotta2017} 	& 81\cite{richert1998}				& $t_\mathrm{w} \sim 4$ months\cite{aliotta2017}\\
SiO$_2$	& * 						&  							& 820-900 			&  								& $\sim$60			&  					& 20\cite{bohmer1993}				& \\  \hline
PHS		& 	45.45				& 							& 33					&  38 							& 	 	 			& 0.40				& $\approx$50						& Particle swap MC \cite{berthier2016prl,berthier2017pnas} \\ 
HS		& 	28.8 					& 27.0						&  					& $^\dagger$						& 69.74 				& 					&								& MD, colloid exp. \cite{taffs2013,royall2015,royall2017}	\\  			
KA 		&  	0.325 				& 0.357(5) 					& 					& $^\dagger$						&  3.62(8)				& 					&								& MD \cite{malins2013fara,royall2015,turci2017prx,coslovichpersonal} \\
Wahn	&  	0.464(7)				& 0.488(5) 					& 					& $^\dagger$						& 1.59(13) 			& 					&								& MD \cite{malins2013jcp,royall2015} \\
\hline
\end{tabular}
\end{adjustbox}
\caption{League table for the bottom of the energy landscape. The equilibrated temperatures $T_\mathrm{eq}$ in this table are collected from the literature specified in each row~\cite{swallen2007,sepulveda2011,zhao2013} or estimated by using enthalpy diagrams~\cite{ramos2011,ramos2015,boucher2011} or estimated the upper limit of $T_\mathrm{eq}$\cite{yu2013} by using the relation between deposition temperature and $T_\mathrm{eq}$ specified in ref. \cite{kearns2007}. 
EB is ethylbenzene,
ECH is ethylcyclohexane,
TNB is 1,3-bis-(1-naphthyl)-5-(2-naphthyl)benzene, 
TL is toluene, 
IMC is indomethacin, 
PR is propene, 
CCl$_4$ is carbon tetrachloride, 
MG is the metallic glass,Cu$_{65}$Zr$_{27.5}$Al$_{7.5}$.
PS is polystyrene,
PCB is polycarbonate, 
AMB is amber, 
OTP is ortho-terphenyl, 
SiO$_2$ is silica,
PHS (very) polydisperse hard spheres,  
HS hard spheres, 
KA denotes the Kob-Andersen mixture, 
Wahn the Wahnstr\"{o}m mixture and 
$t_w$ is the waiting time.
VD denotes vapor deposition, 
MD denotes conventional molecular dynamics simulations. 
Note that, as a strong liquid, the fragility of silica is poorly defined~\cite{angell1995}.
$^\dagger$ No simulations except the particle swaps \cite{berthier2017pnas,berthier2016prl,ninarello2017prx} have yet been equilibrated at deeper supercooling than the experimental glass transition $T_g$. }

\label{tableHowLow}
\end{table*}

Here we review strategies adopted in experiment which can obtain states in the  $T_K<T<T_g$ regime. Our intention is to address the question of how low in the energy landscape experiments have reached. Most of our analysis will consider molecular systems, as these sample phase space \emph{very} much faster than their colloidal counterparts. The latter struggle even to pass the mode-coupling transition, and remain many orders of magnitude in terms of relaxation time from $T_g$ in the definition for molecules, that the relaxation time should be 14 decades more than that of the liquid, $\tau_\alpha(T_g)/\tau_\alpha(T>T_m)\sim10^{14}$.

To assess how close to the bottom of the energy landscape the system has reached, we shall employ the criteria introduced by Ediger and coworkers \cite{swallen2007}.

\begin{equation}
\theta_0(T_\mathrm{eq})=\frac{T_g-T_\mathrm{eq}}{T_g-T_0}
\label{eqThetak}
\end{equation}

\noindent where $T_\mathrm{eq}$ is the temperature at which the system has been equilibrated. Originally $T_K$ is used to represent the ideal glass. However, a value for $T_{K}$ is often harder to obtain than $T_0$. Thus to discuss a wider range of systems from the literature, we recall that $T_0\approx T_K$ for many systems \cite{cavagna2009}. As a result, we consider the extent of traversing between $T_g$ ($\theta_0(T_g)=0$) and $T_0$ ($\theta_0(T_0)=1$).

\subsection{Annealing}
\label{sectionAnnealing}

Implicit in Fig. \ref{figCavagna} is that simply waiting longer than 100 s and reducing the temperature in a commensurate fashion will enable a material to equilibrate at temperatures below $T_g$. In this way, it is conceptually straightforward to equilibrate a material below $T_g$, however the key question is how far below $T_g$ one can reach in a ``reasonable'' waiting time. Inspection of Fig.~\ref{figAngellEternal} indicates that for fragile glassformers, waiting times quickly become very long indeed before the material can be equilibrated. If we take one year as our ``reasonable'' waiting time, then the work of Koh and Simon \cite{koh2013}, who looked at the enthalpy of polystyrene, suggests that with a year of aging one can try to equilibrate to temperatures of 12 K below $T_\mathrm{g}$ or 3\% in absolute terms (see table \ref{tableHowLow}). This means the system can reach 11\% above $T_0$, or $\theta_0=0.23$ in Eq. \ref{eqThetak}. Other examples inlcude the work of Aliotta et al \cite{aliotta2017}, who obtained a fictive temperature of 230K for \emph{ortho}-Terphenyl (OTP). This is consistent with the model of Tool ~\cite{tool1946}, Narayanaswarmy ~\cite{narayanaswarmy1971} and Moynihan~\cite{moynihan1976} (TNM).

The TNM model is based on the assumption that relaxation can be attributed to a \emph{fictive} temperature. Here the term fictive temperature refers to the temperature that an equilibrated material would have, were it to exhibit the same behaviour as the aging material, as determined by some observable, such as the relaxation time. In this case, the fictive temperature is greater than the true temperature of the material, so the relaxation time relevant to the annealing process becomes much faster than the prediction from extrapolation of the VFT expression (Eq. \ref{eqVFT}), which would pertain to the fully aged, equilibrium material. Significantly, the assumption of the TNM model implies the existence of a fast relaxation component.  While other materials with differing fragilities may offer closer approaches to $T_0$ with this conceptually straightforward method, the prediction from the TNM model suggests that without massive increases in waiting time, an approach to around 10\% of $T_\mathrm{K}$ is to be expected.

One way around the limitations of annealing is to use thin films, of polymeric glassformers in particular ~\cite{cangialosia2016}. One recent and exciting result is that it has been shown that thin films of  can exhibit fictive temperatures which approach their Kauzmann temperature, and that further annealing does not ~\cite{boucher2017}.

Before moving on to other strategies to approach the bottom of the energy landscape, we note here an intriguing recent work, which identifies two relaxation timescales in metallic glasses just below their glass transition $T_g$ ~\cite{luo2017}. The additional fast mode of relaxation was suggested by the authors to relate to stress-driven relaxation. It is in fact possible that such behaviour may relate to spinodal gelation ~\cite{royall2018,zaccarelli2007,tanaka2000vps}. Certainly, model glassforming systems are routinely quenched quite close to their gas-liquid demixing line ~\cite{testard2011,testard2014}.

\subsection{Waiting for a long time: Amber}
\label{sectionWaitingAmber}

Another strategy to attain the bottom of the energy landscape is to let nature do the waiting, and have the experiment start some time ago. This approach has been pursued by McKenna and coworkers, who studied fossilized amber, a glass forming biopolymer, which had been aged for \emph{20 million years} \cite{zhao2013}. For amber, ambient temperature (taken as 300 K), is around 100 K below its $T_{g}$ (409 K) and about 40 K below its $T_{0}$ (338 K). As shown in Table \ref{tableHowLow}, the vast timescale of millions of years has taken the fictive temperature to within 8\% of $T_{0}$, the same degree with aging experiments just below $T_{g}$ in the laboratory for one year~\cite{koh2013}, although application of Eq. \ref{eqThetak} shows a more promising $\theta_0=0.61$. It is worth noting here that McKenna and coworkers inferred that the dynamics of amber below $T_g$ should be Arrhenius, which implies that there is no Kauzmann ideal glass transition at all : the material simply becomes gradually slower until absolute zero is reached \cite{zhao2013}. However, this analysis presumes a single characteristic relaxation time. In other polymers ~\cite{cangialosi2013} and metallic glasses ~\cite{luo2017}, multiple relaxation times have been observed upon recovery following aging, suggesting a potentially more complex situation.

\subsection{Ultrastable Glasses}
\label{sectionUltrastable}

\begin{figure*}[!htb]
\centering \includegraphics[width=90mm]{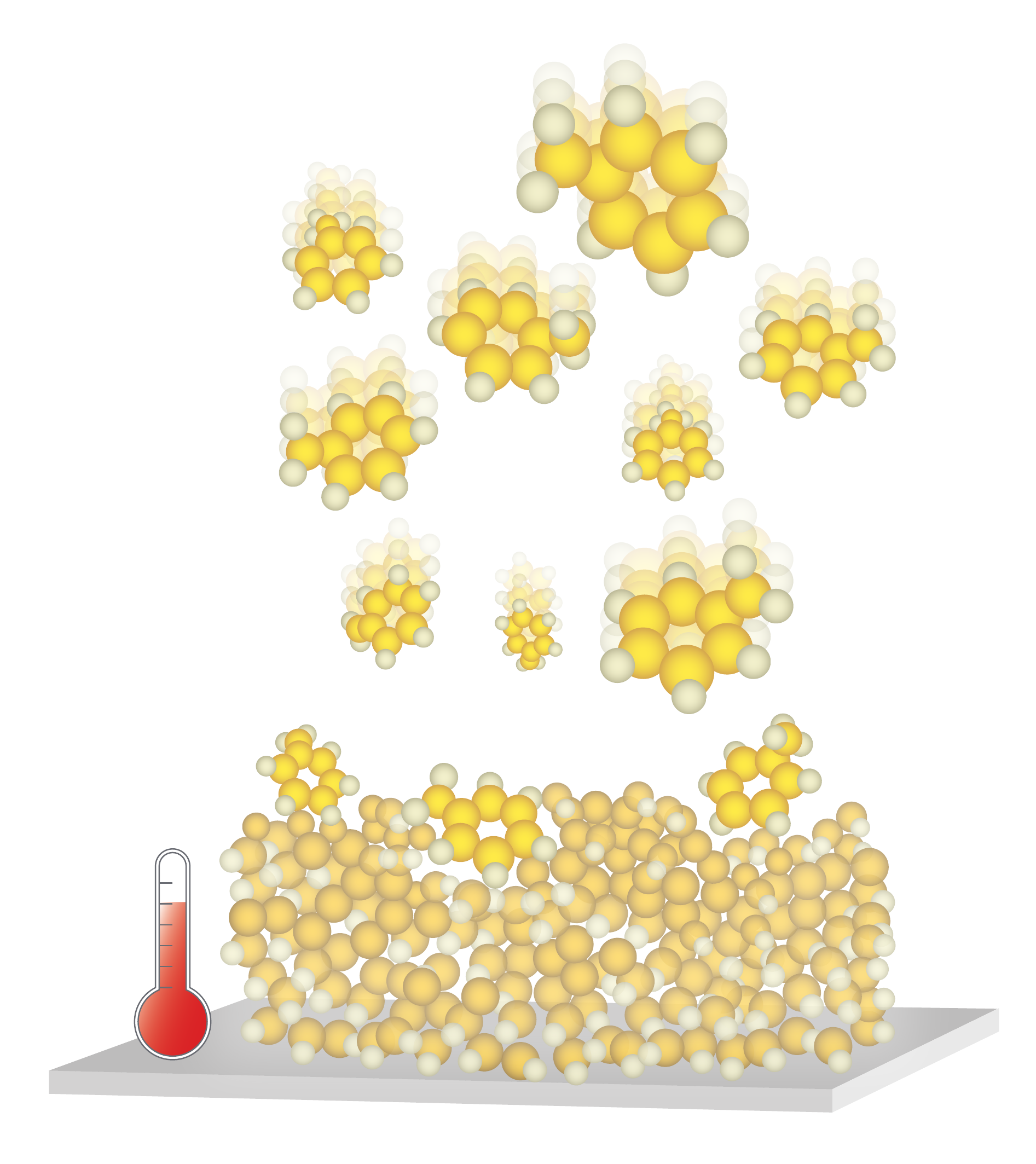} 
\caption{Reaching for the bottom of the energy landscape using free surfaces. Vapor-depositing glass forming molecules where the substrate is at around $0.85 - 0.93 T_g$ leads to \emph{ultrastable glasses}~\cite{ramos2011, ramos2015, swallen2007, sepulveda2011, chua2016, yu2013}. Molecular mobility near the free surface is believed to be the key to producing a such ultrastable glasses. The incoming molecules can reach equilibrium if the substrate is close to $T_g$. Reproduced from \cite{berthier2016pt}.
\label{figBallyBallThermometer} }
\end{figure*}

In a landmark paper \cite{swallen2007}, Ediger and coworkers demonstrated a new and very powerful means to access extreme states towards the bottom of the energy landscape. The method to produce these \emph{ultrastable} glasses is to deposit \emph{e.g,} indomethicin (IMC) or 1,3-bis-(1-naphthyl)-5-(2-naphthyl)benzene (TNB) molecules, one-by-one from a vapor onto a substrate as indicated in Fig. \ref{figBallyBallThermometer}. The depth to which the material can reach in the energy landscape is quantified via its fictive temperature, which is determined from calorimetry and density measurements. The value of the substrate which minimises the fictive temperature is around  0.85 $T_g$. Due not least to their novel material properties, the field of ultrastable glasses has seen remarkably rapid growth, for which we refer the reader to the very recent review by Ediger \cite{ediger2017}.

Subsequent work suggests this technique can lead to conditions which exceed $\theta_0=0.5$, or that the material is equilibrated  to within 10\% of $T_0$ \cite{kearns2008,ramos2011}. Excepting perhaps the intriguing work of McKenna and coworkers (see above) \cite{huang2003, mckenna2015}, this represents the closest approach yet achieved in experiment to $T_\mathrm{K}$. However in the case of a number of molecular glassfomers \cite{lyubimov2015,dalal2015}, there is evidence of anisotropy in the thin films of many ultrastable glasses  \cite{dalal2013,dalal2015}. 
In a sense, when comparing to conventional liquid-cooled glasses, this suggests that there is some fundamental change in the nature of the material. 
Nevertheless, the potential of the vapor deposition technique is further illustrated by its applicability to a diverse range of molecules \cite{ediger2017, chua2015,ramos2015, sepulveda2011, chua2016} some of which, such as toluene ~\cite{rafolsribe2017}, are more isotropic in shape and even to metallic glassformers, in which the interactions are roughly spherical \cite{yu2013}. From the work done on the metallic glassformers by Yu \textit{et al.}~\cite{yu2013}, we estimate an astonishing approach to the ideal glass, in the sense that the parameter $\theta_0=0.7$ or within 7\% of $T_{0}$. The equilibration temperature $T_\mathrm{eq}$ of the system is estimated from extrapolation of the enthalpy in the same manner with other vapor-deposited glass shown in other work \cite{ramos2011,swallen2007}. We note that we could not find an estimation of $T_{0}$ (or $T_K$) in the literature so we employed numerical simulation with an embedded atom model on the same system as shown in Fig.~\ref{figAngellEternal}. We thus estimate $T_{0}$ as $626$ K.

The key to forming these ultrastable glasses appears to lie in the mobility of the surface molecules. Stevenson and Wolynes showed that simple arguments based on Random First Order Transition Theory predict a value of $\theta_0 = 1/2$, due to a halving in the free energy barrier to relaxation in the presence of a free surface. In practise the vapour deposition by which the ultrastable glasses are formed allows for some further equilibration, \emph{i.e.} it appears to be possible to equilibrate to lower temperatures, $\theta_0 > 1/2$  ~\cite{stevenson2008}. These free surface dynamics have been probed in experiment. Using a novel technique specifically designed to probe surface relaxation, by using a tobacco mosaic virus placed on the surface, and using atomic force microscopy to probe the mobility, Fahkraai and coworkers have shown that surface dynamics are decoupled from the bulk and follow an Arrhenius-like scaling \cite{zhang2017,zhang2017prl}. This Arrhenius scaling may enable a  system to get even deeper in the energy landscape than has been possible thus far.

\begin{figure*}[!htb]
\centering \includegraphics[width=90mm]{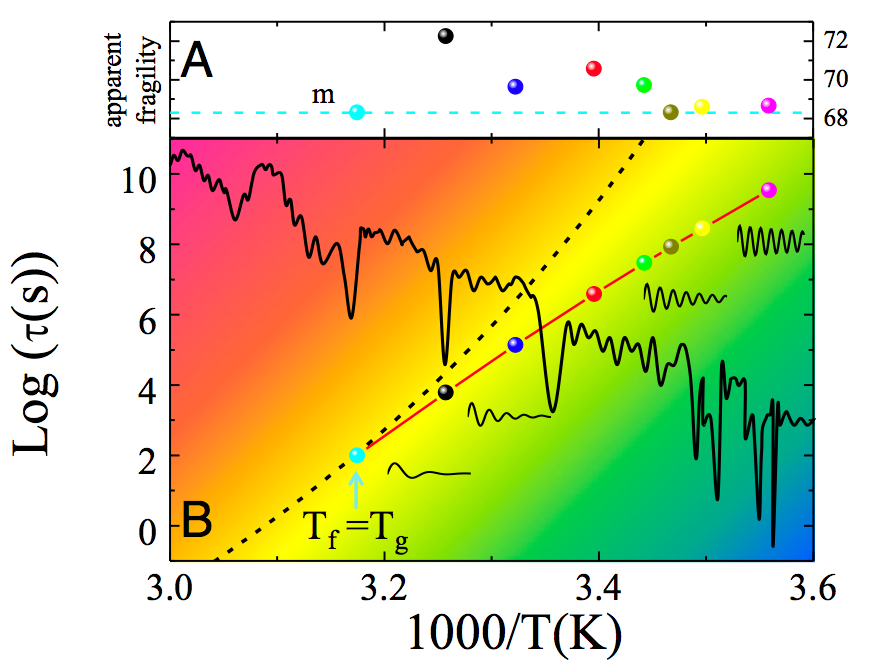} 
\caption{Viscous flow of a supercooled liquid in the exapoise range.
(a) Apparent fragility determined from the sound velocity jump in glasses with different fictive temperatures indicates a fragile to strong transition below T$_g$. Conventional fragility $m =$ is also reported. 
(b) Arrhenius plot for $T < T_g$, obtained by integrating data from (a). A correlation is found between the depth of the inherent structures in the energy landscape (gray line) probed by glasses of different fictive temperature $T_f$ and the mechanical response (sketched by time-domain damped oscillations). The VFT behavior obtained from above $T_g$ data is also shown (dashed line). 
Reproduced from~\cite{pogna2015}.
\label{figSciopignio} }
\end{figure*}

Ultrastable glasses correspond to materials with relaxation times, in the conventional sense, that extend to spectacular timescales. One may nevertheless enquire as to whether there might be some way to deduce a relaxation time, and indeed such an approach has been pursued by investigating the sound velocity in such materials. The results are shown in Fig. \ref{figSciopignio}. Remarkably, these point to an Arrhenius behaviour for the relaxation time -- \emph{i.e. no dynamical divergence until absolute zero and thus no 
transition \paddyspeaks{to an ideal glass}}~\cite{pogna2015}. Interestingly, these results are consistent with those of Mckenna and coworkers who studied hyper-aged amber (section \ref{sectionWaitingAmber}).

One final point to note about ultrastable glasses, suggesting that they may be genuinely novel states, is their tunnelling properties. Quantum tunnelling is typically encountered, and rationalised by a two-state model in conventional liquid-cooled glasses and may be related to small voids in the glass. Ultrastable glasses, on the other hand, do not appear to exhibit such two-state tunnelling \cite{perezcastaneda2014pnas}. Other glassformers which access deep into the energy landscape, such as fossilised Amber, do exhibit tunnelling  \cite{perezcastaneda2014prl}. It is worth noting that hyperquenched glasses (quenched while still quite high in the energy landscape) exhibit enhanced low-frequency relaxations relative to annealed systems \cite{angell2003}.

\subsection{Hypothesis on ultrastable glasses and their relation to fragility}

\begin{figure*}[!htb]
\centering \includegraphics[width=90mm]{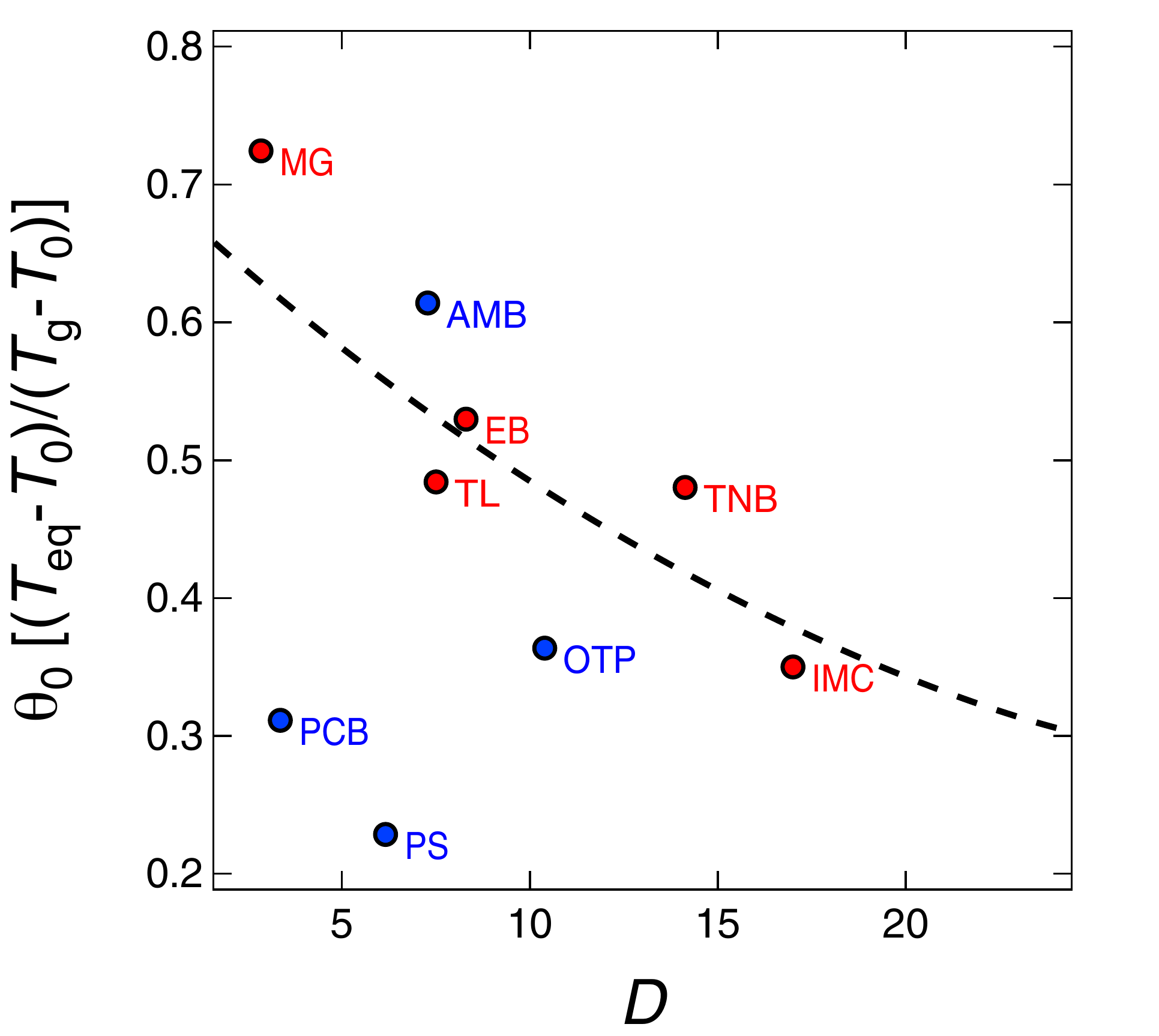} 
\caption{
Approaching the bottom of the energy landscape: the role of fragility.
Here we plot $\theta_0$ vs fragility parameter $D$. The dashed line is a guide to the eyes. Remarkably, a negative correlation is found between $\theta_0$ and fragility parameter $D$ across various substances. Red circles indicate data sets obtained by physical vapour deposition and blue circles indicate data sets obtained by annealing. Abbreviations shown to the right of each plots are the same as in Tab.~\ref{tableHowLow}. 
\label{figSoichi} }
\end{figure*}

As we saw in the previous section, physical vapour deposition at an appropriate temperature allows us to make ultrastable glass even for metallic glassformers, whose constituent atoms interact in an isotropic fashion, as compared to molecular glassformers, which is supported in computer simulation of systems with spherically symmetric interactions (see section \ref{sectionNonLocal}). This suggests that the ability of materials to form ultrastable glasses may be rather general, and so here we provide a discussion concerning relaxation in thin films and provide a connection to \emph{fragility}.

The process of the realization of the vapor-deposited ultrastable glass is pictured as follows; Firstly, highly energized molecules, \emph{i.e.} vapor, attach to the surface whose temperature is somewhat lower than $T_{g}$. Second, only a few molecules on the surface are energized and mobile, which over time relax to a local energy minimum and cease moving (apart from local vibrations).

We noted above in section \ref{sectionDynamicStaticLengthscales} that high fragility implies a larger energy barrier to relaxation and that, within the framework of for example Adam Gibbs \cite{adam1965} or random first-order transition theories \cite{lubchenko2007}, this would be related to a decrease in configurational entropy and thus the cooperatively rearranging regions would increase in size. What does this mean for thin films? We suppose that the influence of the surface in enabling relaxation in the film 
is limited to a lengthscale beyond which it has little effect. Let us further suppose that this surface lengthscale is related to the CRR size. Therefore, if more fragile materials indeed have longer co-operative lengthscales, the size of cooperatively rearranging regions might grow sufficiently that the ability to relax via the surface, crucial for ultrastable glasses, would be lost. Conversely, for strong glassformers, with small co-operative lengthscales, the ability of the surface to enable relaxation may be retained at much deeper supercooling. How might this conjecture be tested? The degree of supercooling made possible by the vapour deposition is characterised by Eq. \ref{eqThetak}. We therefore plot $\theta_0$ as a function of fragility, characterised by the parameter $D$ in Eq. \ref{eqVFT} in Fig. \ref{figSoichi}. Now $D$ measures the inverse fragiity, so we expect a negative correlation, which is precisely what we find for a wide range of materials. Note that in Fig. \ref{figSoichi}, we also plot annealed materials (blue data points), whose scatter is very much larger.  This lack of correlation in bulk (non-vapour-deposited) systems supports the idea of a link between fragility and the effectiveness of vapour deposition in approaching the bottom of the energy landscape.

\subsection{Experiments on Colloids}
\label{sectionColloids}

Before closing the discussion on experimental approaches to access the bottom of the energy landscape, we briefly consider colloids. These systems follow the same rules of equilibrium statistical mechanics as do atoms and molecules, and are sought-after for four properties. Firstly, they can be directly imaged. In particle-resolved studies \cite{ivlev,hunter2012,gasser2009}, their dynamics are tracked at the single-particle level yielding data otherwise available only to simulation ~\cite{crocker1995,ivlev,royall2003,vanblaaderen1995}. Secondly, relative to molecules their large size leads to sluggish dynamics, so time-resolved data can be taken ~\cite{weeks2001,kegel2001} \emph{but} this means that it is hard to access the kind of timescales associated with the molecular glass transition with colloids (Figs. \ref{figCavagna} and \ref{figAngellEternal}). Therefore, for the purposes of probing the depths of the energy landscape, colloids cannot hope to compete with molecules. Thirdly, the interactions between the colloidal particles can be tuned, and only classical interactions need be considered. This means that they form sought-after model systems against which theories based on, for example hard spheres, can be readily compared. Finally, and crucially for our discussion, colloids can be manipulated, in particular by the use of optical tweezers \cite{grier2003,gokhale2016,gokhale2016jsm,williams2015jcp}.

The sluggish dynamics of colloids leads to a common --- and important --- misconception. This is that the relaxation time at the \emph{experimental} glass transition $T_g$ is somehow absolute. Molecules are said to vitrify when the relaxation time exceeds 100~s, which is an entirely anthropocentric quantity. This is around 14 orders of magnitude longer than relaxation timescales in high-temperature liquids. On the other hand colloids can have relaxation times of $\sim100$~s without exhibiting any traces of slow dynamics. This obviously presents a major problem: If we employ for colloids the same criteria as those applied to molecular fluids, we find that the ``molecular'' 100~s corresponds to $\sim10^{8}$~\emph{years}. In other words, in order to presently complete the equivalent \emph{dynamic range} of 14 orders of magnitude in particle-resolved techniques on colloids, the measurement would need to have commenced in the Jurassic period! (and while this is possible with experiments on amber (section \ref{sectionWaitingAmber}), the particle-resolved experiments are in practise limited to a few days duration \cite{royall2014arxiv}). In other words, dinosaurs made little use of particle-resolved studies of colloids.

In terms of colloids reaching for the bottom of the (free) energy landscape, the most deeply supercooled experiments have just passed the mode-coupling transition~\cite{brambilla2009}. As modest as this is in comparison with molecules, the strength of the colloids lies in the quality of information delivered: no other experimental technique delivers 3d time-resolved particle coordinates comparable to computer simulation. In the future we may hope to go more deeply supercooled than has been achieved so far, but even without passing $T_g$ it should nonetheless be possible to discriminate some of the competing theoretical approaches, due to the richness of the data obtained. Furthermore, the degree of manipulation possible means that some techniques from computer simulation may be employed in colloidal systems. One the one hand, these experiments serve as experimental verification of theoretical approaches, and on the other hand, they show that colloidal systems can indeed approach the bottom of the energy landscape more effectively than might be expected from dynamical considerations alone. Considerable progress in this direction has been made through the use of optical tweezers to ``pin'' particles \cite{gokhale2014,gokhale2016jsm,gokhale2016,williams2015jcp}, or spontaneous adsorption in 2d systems \cite{williams2018} (see simulation work in Section \ref{sectionPinning}). Meanwhile insight can be gleaned through ingenious experiments on ``vapour-deposited'' colloidal glasses \cite{cao2017}.

\begin{figure*}[!t]
\centering \includegraphics[width=70mm]{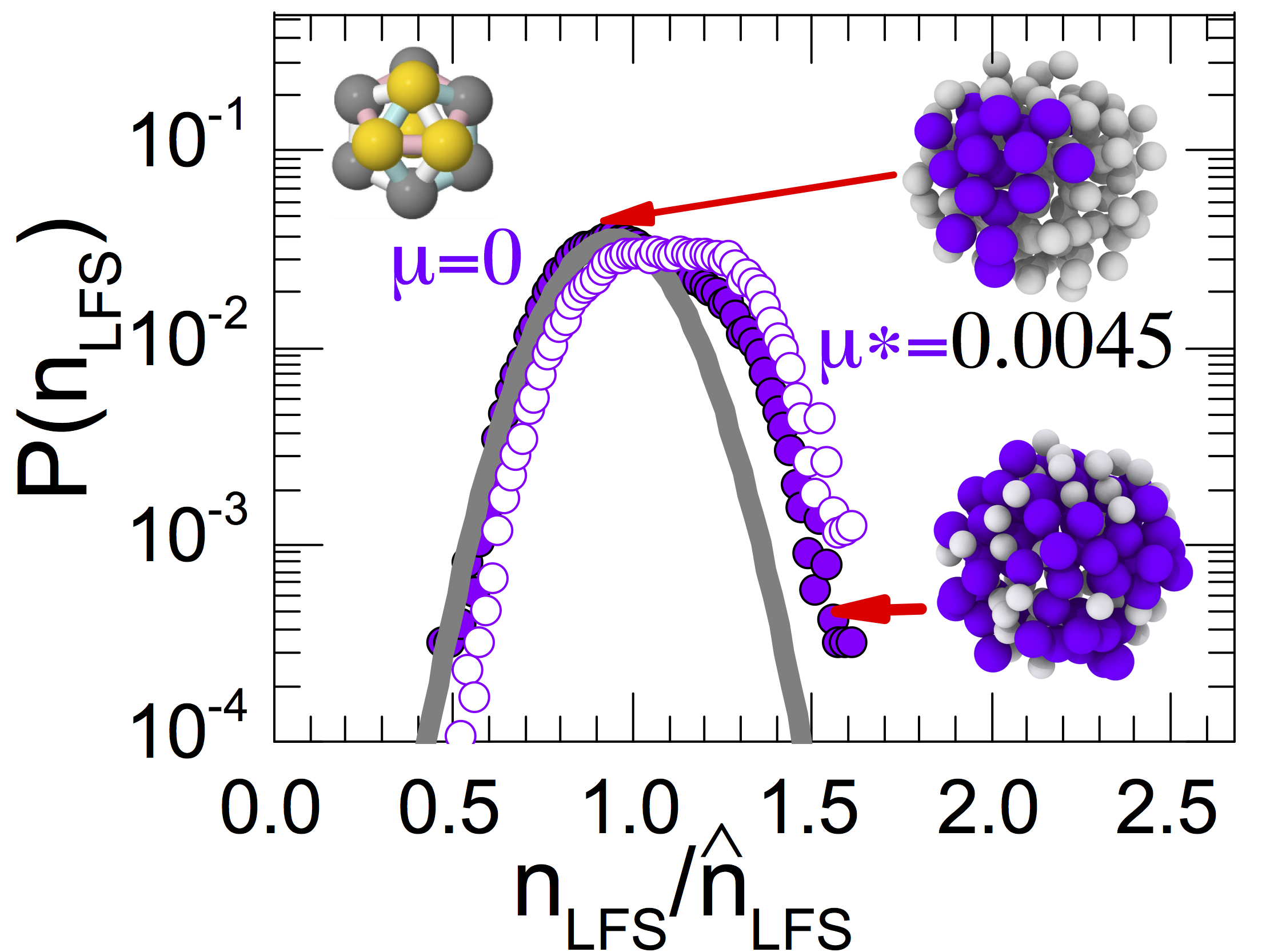} 
\caption{
Probability distributions of populations (filled symbols) of particles in locally favoured structures in trajectories sampled from experimental data. The locally favoured structure is the defective icosahedron \cite{royall2017}, indicated top left. Data is then reweighted according to a dynamical chemical potential $\mu$ (open symbols) to reveal a phase transition at $\mu^*$ between a local-structure-rich state (right) and a normal liquid. In the snapshots of the experimental coordinates, purple particles are in defective icosahedra, rendered for each state. The population of purple LFS particles is much higher in the LFS-rich state, indicating that it lies deep in the energy landscape. Reproduced from~
\cite{pinchaipat2017}.
\label{figRattachai} }
\end{figure*}

Even with current technology, it is possible to approach the bottom of the energy landscape more closely that the arguments above suggest. Using methods developed in computer simulation \cite{speck2012} (see section \ref{sectionDynamicalPhaseTransitions}), Pinchaipat \emph{et al.} \cite{pinchaipat2017} were able to identify a non-equilibrium phase transition. States identified in this non-equilibrium phase transition correspond to configurations very deep in the energy landscape. Under the assumption that the number of locally favoured structures (geometric motifs associated with supercooling \cite{royall2015physrep}) represents the degree of supercooling, the configuration illustrated in Fig. \ref{figRattachai} is at an effective volume fraction $\phi_\mathrm{eff}=0.59$, beyond that accessible to conventional techniques. While still not as deeply supercooled as the molecular glass transition, nevertheless this corresponds to a relaxation time around 100 times higher than the typical limit of particle-resolved colloid experiments.

\section{Approaching the bottom of the energy landscape \emph{in silico}}
\label{sectionSimulation}

Computer simulations of deeply supercooled liquids encounter difficulties analogous to those that afflict experiments: if both the structure and the dynamics of the liquid are to be directly sampled, at low temperatures (or very high packing fractions in the case of hard-sphere-like systems) the very slow glassy dynamics makes brute force calculations extremely demanding in terms of computational resources. Around the (avoided) dynamical transition temperature $T_\mathrm{mct}$ predicted by Mode Coupling Theory, such brute force calculations rapidly become impossible even for the simplest models (such as polydisperse mixtures of hard spheres or binary mixtures of Lennard-Jones particles) in the sense that (metastable) equilibrium is never attained. The liquid slowly and continuously drifts to lower (free)-energies, undergoing ageing on the accessible simulation timescales.

However, an important advantage of theoretical and numerical approaches with respect to experiments is that it is possible to perform controlled approximations that allow us to circumvent the contingent limits of our computational power. Such approximations rely on a limited number of assumptions, and aim at sampling equilibrium properties, artificially accelerating the local relaxation processes by different means.

Here we summarise some of the most popular and intriguing ideas that have been developed in the last decades to explore deeply supercooled, low (free) energy states: we will first discuss evaluating the energy landscape itself, and then consider  highly optimised, non-local Monte-Carlo schemes allowing to rapidly sample the equilibrium configurations of deeply supercooled liquids (event-chain algorithm; particle swapping techniques); we will then consider the route to very large static correlations obtained through \textit{particle pinning}, inspired by Random First Order Transition theory (RFOT); then, the numerical equivalent of the experimental vapour-deposition techniques will be illustrated, followed by a discussion of trajectory-based techniques that (together with non-equilibrium information) allow us to access very low energy states from the statistics of rare events; we conclude with direct methods of local minimisation and sampling of the potential energy surface.  Before exploring the what may loosely be termed the ``smart algorithm approach'', \emph{i.e.} the techniques such as those mentioned above, we note that with the advent of code optimised for graphic processing units (GPU), it has now become possible to run brute force simulations for considerably longer than has been possible until very recently: for example, with GPU-codes highly optimised for systems with a few thousands of particles, such as RUMD \cite{bailey2017rumd}, or with massively parallel codes such as LAMMPS \cite{plimpton1995fast,brown2011} or HOOMD-blue \cite{anderson2008,glaser2015}, which can scale to thousands of GPUs. Nevertheless, in terms of the approach to the bottom of the energy landscape, even such GPU optimised code appears to be outperformed by some highly specialised algorithms.

\subsection{Evaluating the energy landscape}
\label{sectionEvaluatingTheEnergyLandscape}

In recent years, improvements in computational processing power have enabled development of techniques to investigate some of the predictions of the energy landscape concept  ~\cite{goldstein1969,sciortino2005}. Among the first developments ~\cite{stillinger1983,stillinger1984} was to establish the \emph{inherent structure}. The inherent structure is the result of a steepest-descent quench to $T=0$, which thus takes the system to its nearest local minimum in the potential energy landscape (Fig. \ref{figEnergyLandscape}). Stillinger and Weber ~\cite{stillinger1983,stillinger1984} further showed that for low temperatures (where the system is confined to a particular basin for long periods), the entropy could be expressed as a sum of a configurational and vibrational part, $S=S_{\mathrm{conf}}+S_{\mathrm{vib}}$. The vibrational part can be treated harmonically, and combined with the inherent state contribution $S_\mathrm{conf}$, it is possible to ``numerically evaluate the energy landscape'' ~\cite{heuer2008,sciortino1999,karmakar2009,sastry2001,sastry1998,schroder2000}.

Such investigations have largely confirmed Goldstein's energy landscape picture, that the system can be viewed as residing in basins for times of order of the structural relaxation time before undergoing rapid transitions between basins in the energy landscape ~\cite{goldstein1969}. In particular, Doliwa and Heuer ~\cite{doliwa2003} have investigated local structure in the basins (via the potential energy). Basins with high activation energy barriers for escape corresponded to low potential energy. They found that escape from a particular basin corresponds to a complex multistep process involving a succession of energy barriers. Interestingly from the point of view of local structure, some work has suggested that such transitions involve small compact clusters (which might correspond to co-operatively re-arranging regions) \cite{appignanesi2006}, unlike the string-like motion identified in earlier studies ~\cite{schroder2000,donati1998}.

\begin{figure}[!htb]
\centering \includegraphics[width=60mm]{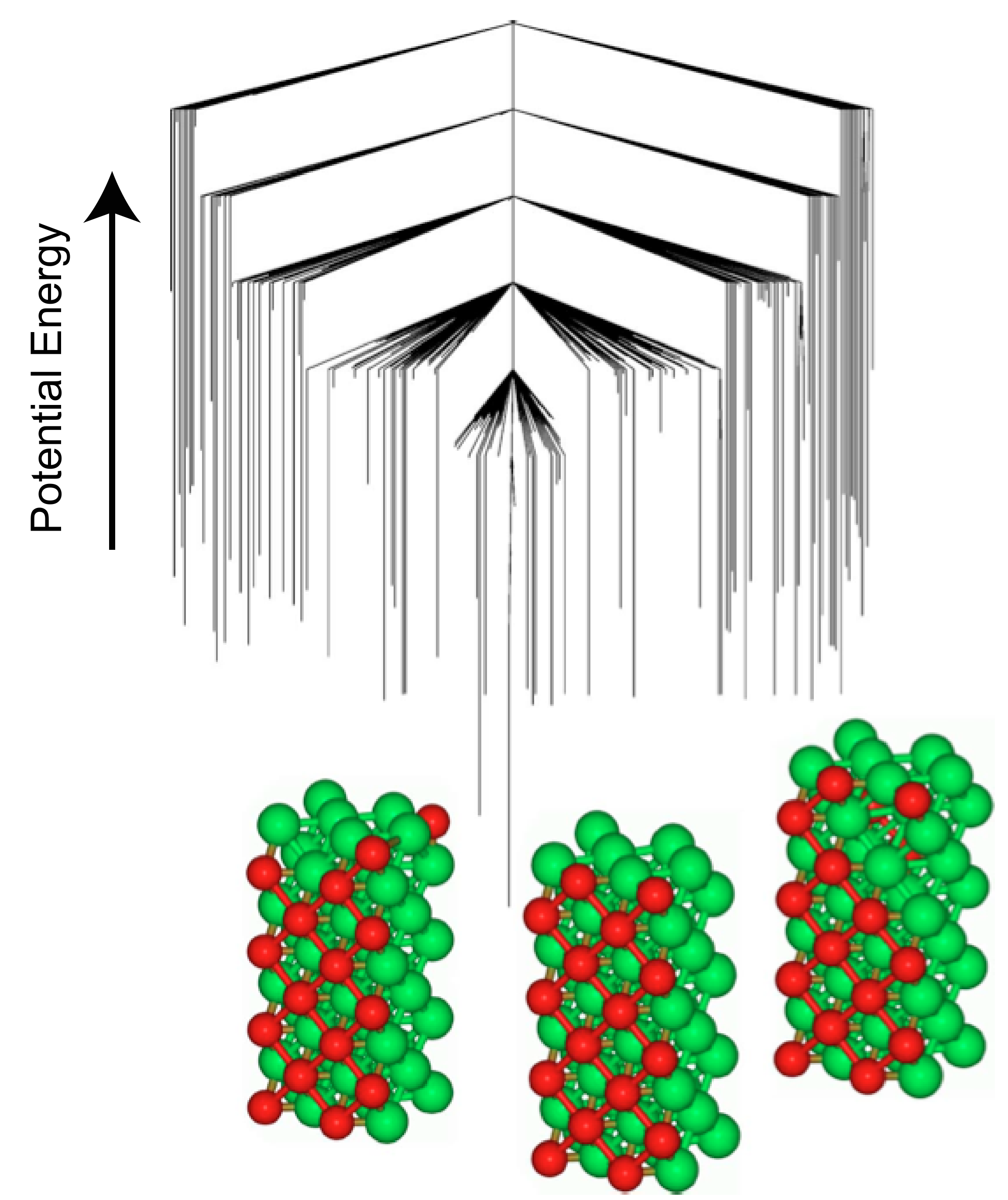} 
\caption{The energy landscape of a $N=320$ Binary Lennard-Jones (Kob-Andersen) system represented as a disconnectivity graph. Three low-lying minima which have a crystal-like appearence are rendered ~\cite{calvo2007,middleton2001}. Note the similarity of these crystals to the bicapped square antiprism which is known to be locally favoured structure for this system ~\cite{coslovich2007,malins2013fara}. Reproduced with permission from American Institute of Physics Copyright 2007.
\label{figVanessa} }
\end{figure}

Wales and coworkers have developed powerful numerical techniques to enable the energy landscape of systems with a few hundred particles to be explicitly calculated ~\cite{wales}. This is sufficient to explore glassy behaviour, such that the energy of a representative set of basins in the energy landscape can be expressed in a ``disconnectivity graph'' (Fig. ~\ref{figVanessa}) ~\cite{calvo2007,desouza2008}. Such analysis yields the minimum energy structure, which, in the case of the Kob-Andersen system appears to be crystalline ~\cite{middleton2001} with a striking structural similarity to the bicapped square antiprisms which are the locally favoured structure for this model~\cite{coslovich2007,malins2013fara}.

\subsection{Non-local Monte-Carlo techniques}
\label{sectionNonLocal}

Sufficiently supercooled liquids are characterised by the slow motion of particles, often \textit{caged} by their neighbours, in a process that dramatically reduces the diffusivity. This phenomenology is well reproduced by thermalised Molecular Dynamics (MD) or local Monte Carlo (MC) schemes, akin to over-damped dynamics \cite{flenner2015}. However, caging prevents full equilibration at very low temperatures. It is therefore desirable to construct dynamics different from the real dynamics that can, however, attain the correct equilibrium distribution of states. Several alternative schemes are based on extended ensemble techniques~\cite{frenkel}, such as replica exchange Markov chain Monte-Carlo (also known as parallel tempering) \cite{swendsen1986replica,hukushima1996exchange,sugita1999replica,earl2005parallel} eventually combined with importance sampling methods such as umbrella sampling \cite{kumar1992weighted,bartels2000analyzing}. In these methods, the dynamics is modified in the sense that several copies of a glassy system run in parallel and exchange information regularly, so that tunnelling across metastable states is possible in order to obtain well equilibrated samples at low temperatures. In these schemes, the microscopic dynamics typically remains physical (for example Newtonian or Langevin dynamics) and glassy phenomena such as caging are still preserved, causing the unavoiidable emergence of dynamical slowing down at low temperatures.

\begin{figure}[t]
\centering
\includegraphics{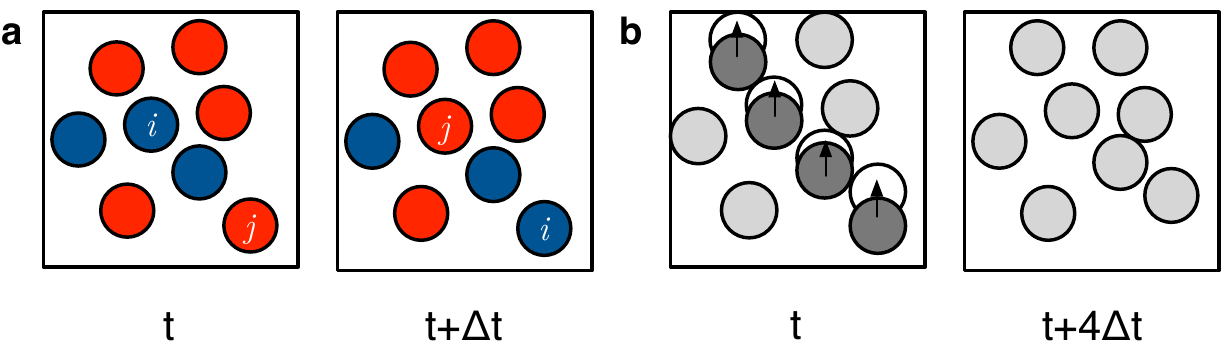}
\caption{Non-local Monte-Carlo schemes. (a) Particle swap in binary mixture of red and blue particles. \cite{berthier2016prl,ninarello2017prx} (b) Straight event-chain Monte-Carlo move: a random direction (in this case along the vertical direction) is chosen and a chain of moves is triggered starting from the particles in the bottom left and involving four particles (dark shading) ~\cite{bernard2011,bernard2009}.}
\label{figMCschemes}
\end{figure}

One can overcome the issue of caging in a more drastic way if non-local moves of the particles are introduced. These allow faster equilibration time, at the cost of loosing the information on the real dynamics of the system. Particularly successful techniques include the following methods.

\paragraph{Particle Swaps Monte-Carlo } Model glassformers are typically mixtures of multiple (at least two) components of different sizes (and sometimes masses). It is possible to exploit the difference in species extending the Metropolis-Monte-Carlo algorithm with moves in a space of particle identifiers, stochastically accepting/rejecting \textit{swaps} of the particle types \cite{gazzillo1989} [see Fig.~\ref{figMCschemes}(a)]. This allows for a new kind of particle move, which contributes to a much more rapid exploration of the configurational space. In practice, while in local Monte-Carlo schemes cage-breaks are a cooperative process, particle swaps make them non-cooperative. Particle swapping techniques have been shown to outperform methods such as parallel tempering for soft binary mixtures \cite{grigera2001}, and very polydisperse mixtures of hard spheres \cite{berthier2016prl}. In general, the efficiency of the method depends on the specific interactions and the composition of the liquids under consideration. In the case of binary mixtures, for example, the efficiency of the method can be affected by undesirable crystallization events (additive mixtures with not too large size ratios) or by very large rejection probabilities (non-additive mixtures such as the Kob-Andersen mixture) \cite{hocky2012,ninarello2017prx}. A careful choice of the size distribution (for example, using \textit{hybrid} mixtures with two main components and a residual component with a broad distribution of possible sizes) allows one to obtain equilibration times that are several orders of magnitude faster than conventional (local) Monte-Carlo methods \cite{ninarello2017prx}.

Such an acceleration of the dynamics allows us to explore a much broader range of temperatures and access equilibrated states with very low configurational entropy 
\paddyspeaks{at finite (reduced) pressure, corresponding to an ideal glass transition} \cite{berthier2017pnas}, \paddyspeaks{while in 2d such a transition would appear to be found only at zero temperature \cite{berthier20182d}}.

\begin{figure}[hbt]
\centering
  \includegraphics[width=0.75\textwidth]{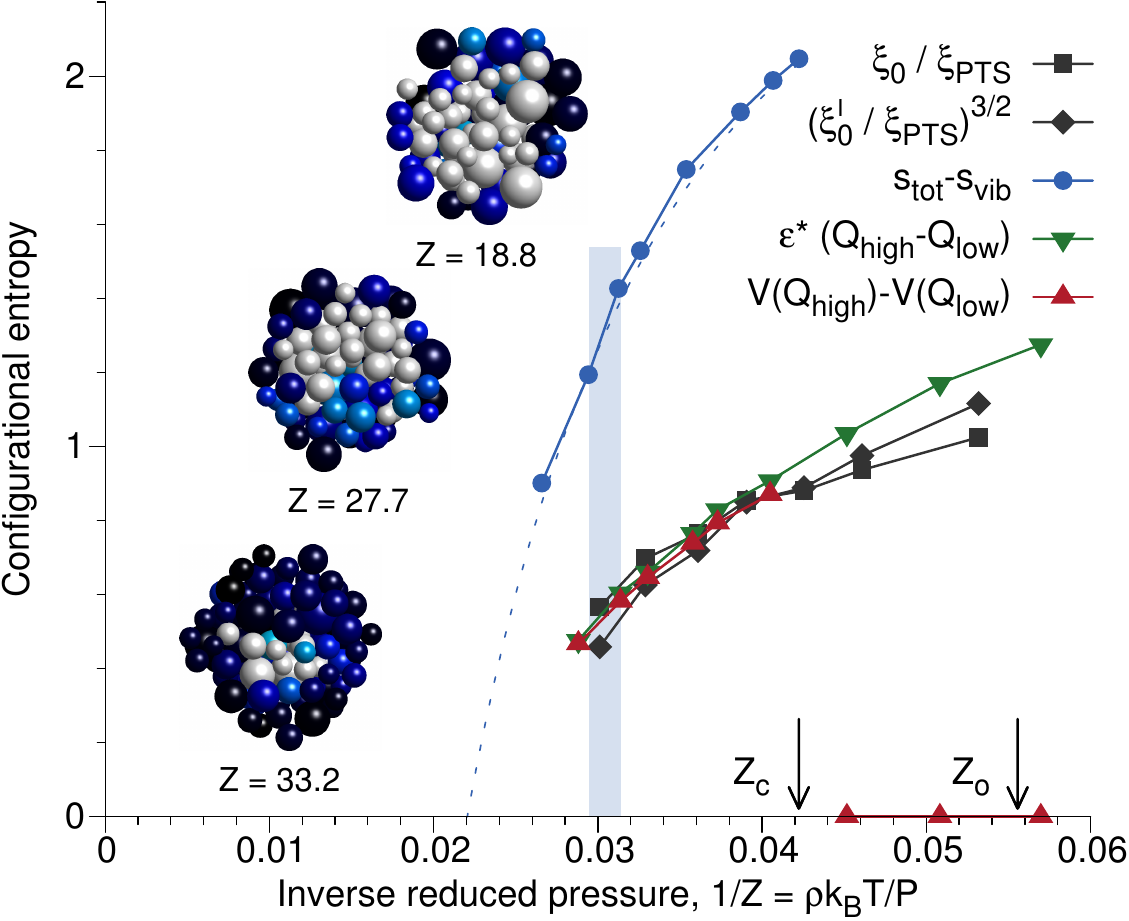}
  \caption{Configurational entropy in a system of polydisperse hard spheres measured according to several indicators (scaled point-to-set length $\xi_\mathrm{PTS}$, configurational entropy from thermodynamic integration $S\mathrm{tot}-S_\mathrm{vib}$, from differences in the Franz-Parisi potential on overlap fluctuations $V(Q_\mathrm{high})-V(Q_\mathrm{low})$, from the susceptibility in biased simulations $\varepsilon^* (Q_\mathrm{high}-Q_\mathrm{low})$). The inset illustrates overlap profiles in cavities. Reproduced from \cite{berthier2017pnas}.}
  \label{figSho}
\end{figure}

Assuming that these methods sample the same configuration space as conventional techniques, these novel results provide ground for a revived debate on the role of multi-body correlations in the emergence of slow dynamics. If, on the one hand, we have clear evidence for the growth of static correlations and the reduction of configurational entropy (regardless of its specific definition, as demonstrated in \cite{berthier2017pnas}, see Fig. \ref{figSho}), on the other hand the simple fact that swap methods push the dramatic slowing down to much lower temperatures suggests that the \textit{observed} dynamic arrest in real systems (which certainly cannot undergo swap moves) is mostly due to the slowdown of local dynamics \cite{wyart2017}, \paddyspeaks{although this suggestion has been challenged with the idea of \emph{crumbling metastability}
\cite{berthier2018}, and a mode-coupling theory based approach similarly leads to dynamical arrest which is suppressed but essentially unaltered particle swaps \cite{szamel2018}. Nevertheless, the idea of a dominant local dynamics of Wyart and Cates  \cite{wyart2017}} 
appears to be corroborated by recent mean-field replica liquid  \cite{ikeda2017} \paddyspeaks{and effective potential \cite{brito2018}} calculations that account for the effect of swap dynamics, suggesting that for the extreme case of such polydisperse systems the slowing down would not be related to the actual metastable states but to kinetic effects.

The particle swaps enable a powerful tool to test ideas which have been previously inaccessible. Examples include demonstrated that surface mobility is indeed the mechanism for the deeply supercooled ultrastable glasses (see section \ref{sectionUltrastable}) \cite{berthier2017prl} and provide evidence in support of the non-equliibrium Gardner transition between glasses and jammed-like states \cite{charbonneau2014,jin2017}.

\paragraph{Event-Chain Monte-Carlo } An alternative route consists of implementing Monte-Carlo schemes that accelerate the equilibration through long-range coherent motions of the particles~\cite{bernard2011,bernard2009}. Such so-called \textit{event-chain} Monte-Carlo algorithms efficiently sample equilibrium configurations minimising~\cite{jaster1999} or avoiding~\cite{bernard2009} rejections, and are therefore particularly suitable for dense systems. In fact, event-chain algorithms have been used to successfully assess the first-order nature of the transition from liquid to hexatic order in two-dimensional hard disks~\cite{bernard2011}, and recently applied to binary mixtures of hard disks~\cite{russo2015,russo2017disappearance}.

In its simplest form (for example, hard-core interactions), the method consists in a deterministic chain of events occurring in randomly distributed directions: a random particle and a random direction are initially selected; the particle moves till it hits a second particle, which moves in the same direction till it hits a third particle, and so on until an arbitrary total displacement is performed [see Fig.~\ref{figMCschemes}(b)]. Such moves are repeated in random directions and ensure that only valid configurations are sampled. The method can be adapted to break detailed balance~\cite{diaconis2000} while still preserving the correct probability measure~\cite{bernard2009} (leading to a further performance boost) and also to work with continuous interaction potentials~\cite{bernard2012addendum,michel2014,peters2012}.

\subsection{Pinned particle techniques} 
\label{sectionPinning}

At very low temperatures, the mobility of the particles of a super-cooled liquid is much reduced, with a highly heterogeneous spatial distribution. The size of the least mobile regions increases as the temperature is reduced, and a so-called \textit{dynamical length scale}, increasing by a factor of $5-10$, is associated with this process in many different glass forming systems \cite{ediger2000,malins2013jcp,berthier2007i,berthier2007ii,flenner2011,franz2000,harrowell2011,karmakar2014,rotman2010,scheidler2004}. The definition of an analogous \textit{static length scale}, growing as the system is cooled, has proven to be challenging \cite{dunleavy2015,royall2015physrep,coslovich2007}, due to the modest impact that the supercooling has on immediate measures of static correlations such as the radial distribution functions, and to observe a well-coupled growth of structural and dynamic lengthscales one needs to go to the limit of weak geometric frustration \cite{turci2017prl}, not suitable for good glass formers. Higher order correlation functions have therefore been developed and so-called point-to-set (PTS) correlations have been defined as multi-point correlations between a probe particle and a reference ``set'' of particles \cite{bouchaud2004,montanari2006,franz2011}. In simulations, the reference set of particles is normally realised through the freezing (\textit{pinning}) of a suitable collection of particles, selected from an \textit{equilibrated} configuration of the system \cite{cavagna2007}. The probability to find the probe particle at a certain relative position with respect to the set is then evaluated, and a corresponding static length $\xi_\mathrm{PTS}$ can be defined \cite{biroli2008}.

\begin{figure}[t]
\centering\includegraphics{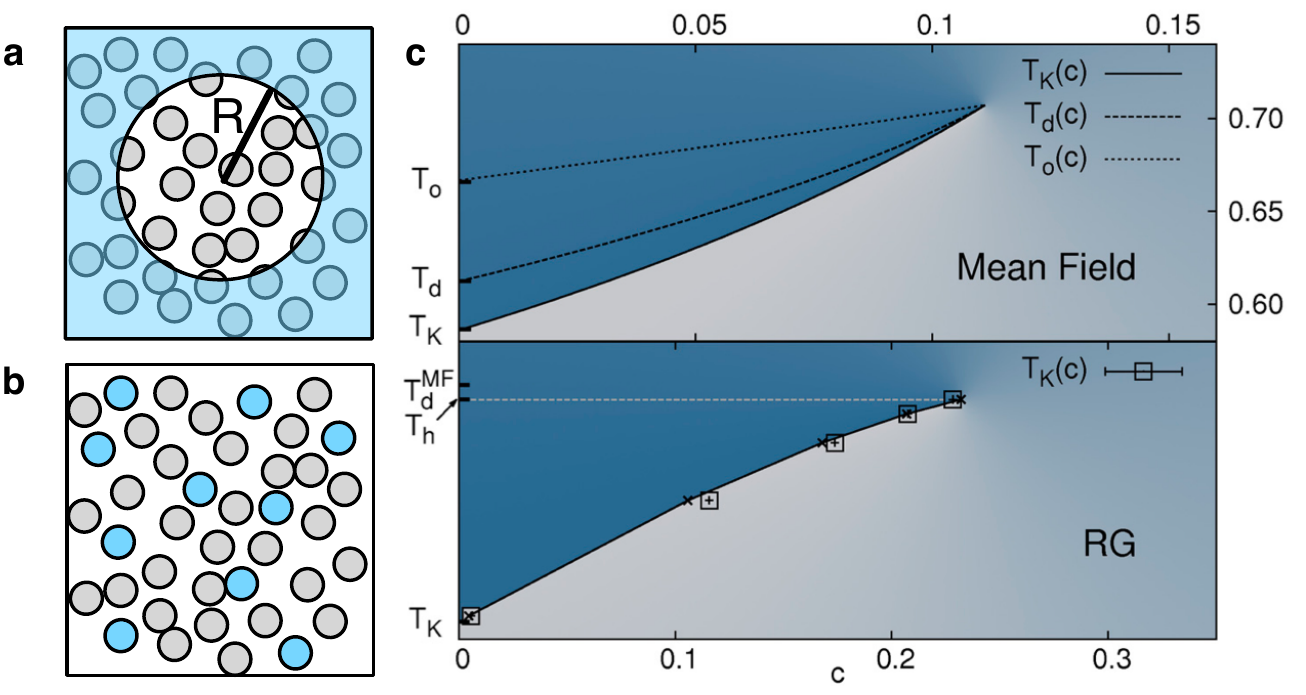}
\caption{Pinning techniques. 
(a) Particles outside a region of radius $R$ are frozen in cavity pinning. 
In random pinning (b) an arbitrary concentration $c$ of particles is pinned (light blue) from an equilibrium configuration. Reproduced with permission from \emph{J. Chem. Phys.} \textbf{131}, 194901 (2008). Copyright 2008 American Chemical Society.
(c) Temperature vs pinning concentration $c$ phase diagram in the p-spin model from mean field (top) and renormalization group calculations (bottom): marked are the glassy dynamics onset temperature $T_o(c)$, the Mode-Coupling critical temperature $T_d(c)$, the ideal glass transition temperature $T_K(c)$. The phase diagram presents a coexistence line (continuous curve) distinguishing a high overlap form a low overlap region. The line starts at $T_K$ for $c=0$ and terminates at a critical point at $T_h$. Reproduced from \cite{cammarota2012}.
} 
\label{figPinningRGMF}
\end{figure}

Even if the method does not allow us to equilibrate the system more rapidly at lower temperatures, it has been shown that via the pinning of some sets of particles it is possible to easily access a critical-like regime at relatively high temperatures. In this regime, the degree of similarity between different configurations (as measured by the so-called \textit{overlap} order parameter) fluctuates and a first-order, equilibrium phase transition between configurations with low and high overlap can be detected. Within the framework of Random First Order Transition theory (RFOT) \cite{lubchenko2007}, this transition can be seen as reminiscent of a low-temperature, inaccessible, thermodynamic transition from the supercooled-liquid to an ``ideal'' glass, with configurational entropy comparable to that of a crystal, but devoid of long-range order. In this sense, pinning would allow us to access, at higher temperatures, a high-overlap phase which would be representative of the kind of amorphous phase associated with the ``ideal'' glass phase. It is implicit to observe that such a high overlap phase is also a phase with very low mobility.

In practice, two methods have been followed in order to measure the point-to-set correlation lengths $\xi_\mathrm{PTS}$:

\textit{Cavity pinning. --- } In an equilibrated liquid configuration a cavity of a given radius $R$ can be identified, outside which all particles are pinned. The pinning of the particles on the edge of the cavity reduces the configurational entropy of those inside, biasing the system in favour of arrest. If the radius $R$ of the cavity is larger than $\xi_\mathrm{PTS}$, the particles at the centre remain mobile, if the cavity is smaller than $\xi_\mathrm{PTS}$, the entire cavity is immobilised (Fig. \ref{figSho} insets)\cite{biroli2008,charbonneau2016,cammarota2009}.

\textit{Random pinning --- } Rather than pinning all particles outside a cavity, one can instead pin a certain proportion throughout the system \cite{berthier2012,karmakar2013}. The concentration of this pinned population of particles $c$ leads to a mean separation $\ell_c=c^{-1/d}$ (where $d$ is dimension). When $\ell_c<\xi_\mathrm{PTS}$, it is argued that an ideal glass transition corresponding to that predicted by RFOT around $T_K$ is found \cite{cammarota2012,cammarota2013,kob2013}. It is worth noting that pinned configurations are (with sufficient sampling) formally structurally indistinguishable from their unpinned counterparts at the same temperature\footnote{Crudely speaking this can be seen from the fact that the pinned system interacts via the same Hamiltonian as does the unpinned system, and that pinned particles are chosen to be frozen in an equilibrium supercooled liquid \cite{jack2014}.}. Nevertheless increasing the concentration of pinned particles means that the configurational entropy drops and strong many-body correlations are established \cite{fullerton2014}. Recently, experimental evidence has been found for dynamical arrest upon pinning using colloidal suspensions (see Section \ref{sectionColloids}) \cite{gokhale2014,gokhale2016jsm,gokhale2016}.

\subsection{Simulated vapor deposition}
\label{sectionSimVapDep}

Via Molecular Dynamics and kinetic Monte-Carlo techniques it is possible to reproduce \textit{in silico} the physical process of vapor deposition of simple model systems. These methods have originally been developed to study phenomena such as condensation on surfaces~\cite{abraham1970}, epitaxial growth~\cite{schneider1985} or the properties of thin films such as diamond~\cite{battaile2002,battaile1997} and graphene~\cite{meng2012} from the material science point of view.

Recently, prompted by the achievements in experimentally deposited glasses (see section \ref{sectionUltrastable}), these methods have been applied to produce ultrastable states for model glass formers. The simulation protocol is designed to reproduce the physics of experiments: a substrate is thermalised separately from vapor molecules that are gradually deposited at random locations and orientations over it; after energy minimization and a short equilibration run, the procedure is reiterated so that a thick film of deposited particles can be grown.

Glasses prepared through vapor deposition acquire a higher kinetic stability due to the higher density that they manage to achieve \cite{qi2016,singh2011,singh2013,singh2014}. In simulations of polymeric glasses, it has been shown~\cite{lin2014} that while such increased stability is correlated with a high degree of anisotropy in the final structure of glasses formed by long polymeric chains (with evident layering in the direction orthogonal to the substrate), short polymer chains still access very low energy regions of the energy landscape and remain isotropic. Results, for small molecules, which relate the nature of the orientation to the substrate temperature, reproduce experimental results of anisotropic glasses in the form of laterally oriented molecules for very low temperatures and vertically oriented molecules for temperatures mildly below the glass transition where ultrastable glasses teaches their optimal fictive temperature~\cite{lyubimov2015}.

For models with spherically symmetric interactions, vapor deposition is equally successful and generates isotropic glasses that are as low in energy as glasses obtained from cooling rates that are 2-3 orders of magnitude lower than those used in ordinary simulations of glasses~\cite{lyubimov2013}.  While the structures  that  are formed in vapour-deposited systems are statistically identical to those observed in ordinary glasses compared at the same inherent state energy ~\cite{reid2016}, compositional inhomogeneities can be measured at the interfaces of the films formed by vapor deposition \cite{singh2014,reid2016}. Recently, computer simulations of a polydisperse Lennard-Jones model \cite{berthier2017create} provided further evidence for the idea that the increased kinetic stability as a consequence of the enhanced surface diffusion of vapor deposited glasses, suggesting a possible explanation of the link between the experimentally observed surface mobility \cite{swallen2007,zhu2011} and the properties of these ultrastable glasses.

\subsection{Large deviations of time-integrated observables}
\label{sectionDynamicalPhaseTransitions}

As previously mentioned, very low temperature (and low free energy), supercooled liquids are characterised by strong dynamical heterogeneities~\cite{berthier}. In particular, dynamical arrest corresponds to the growth of large regions where the particle motion is suppressed and only rare thermally activated processes succeed in triggering some mobility.

It is possible to provide a quantitative spatio-temporal description of dynamical heterogeneities with the use of space and time integrated physical observables, defined over realisations or \textit{trajectories} (also called \textit{histories}) of a system. Within the framework of large deviation theory applied to non-equilibrium systems~\cite{bodineau2004,elmatad2010,derrida2001,touchette2009}, free energies for trajectory-dependent observables have been defined, which led to the determination of \textit{phase transitions} in trajectory space for model systems of glassy dynamics~\cite{chandler2010,speck2012,garrahan2007,hedges2009,merolle2005,speck2012jcp}. Such phase transitions depend on the choice of the time-extensive observable under consideration: oftentimes, it is the activity (\textit{i.e.} the particle mobility) that it is considered~\cite{hedges2009}; for systems under external forces such as shear, the total particle current has been studied ~\cite{turci2011}; when structural information is at hand, it is also possible to compute time integrals of the concentration of well-defined structural motifs and arrangements~\cite{speck2012}. In all cases, it appears that, for temperatures below the onset temperature, schematic and atomistic models of glassy dynamics present the signature of a \textit{dynamical phase coexistence} between trajectories with high mobility/high currents/low concentration of structures and low mobility/low currents/high concentration of structure. While the former characterises high energy liquid states, the latter can be seen as representative of low temperature arrested states.

The approach in terms of large deviations of time-integrated observables is inherently dynamical. Connecting such a dynamical description to the thermodynamics of supercooled liquids is a challenging and ongoing research direction~\cite{coslovich2016jstat}. In the following we discuss in more detail two complementary approaches to explored dynamical phase transition in glassy systems: (a) the measurement of activity (\textit{i.e.} time integrated mobility) and its fluctuations in the so-called \textit{s-ensemble}; (b) the measurement of the time-integrated density of locally favoured structures and its fluctuations, the so-called \textit{$\mu$-ensemble}.

\paragraph{The \textit{s}-ensemble}
In the context of dynamical arrest, dynamical phase transitions have been first identified in simple models with trivial energetics but non-trivial dynamical rules that mimic steric hindrance and jamming, \textit{i.e.} kinetically constrained models~\cite{ritort2003}. Chandler and Garrahan and coworkers \cite{garrahan2007,merolle2005} tracked the activity per trajectory (total particle dispacements or spin flips) of particles (or spins) on one and two dimensional lattices and computed its probability distribution through the generation of many trajectories of the system \cite{bolhuis2002, giardina2006}. At low enough temperatures, the probability distribution shows non-gaussian tails, with slow, \textit{inactive} trajectories being over-represented.  Such ``fat tails'' in the distribution are akin to a thermodynamic transition in a small system (say the liquid-vapour transition) where the system samples configurations from both phases with some probability and where one controls temperature to approach phase transitions and critical points. Within the framework of large deviation theory one can show that the formal analog of the thermodynamic temperature is the conjugated field $s$. In general terms, it is possible to write that the partition function $Z_{C_\tau}(s)$ for trajectories of length $\tau$ characterised by the total activity $C_\tau$ can be obtained from the probability distribution as
\begin{equation}
Z_{C_{\tau}}(s)=\sum_{\rm trajectories} {\rm Prob(trajectory)}e^{-s C_{\tau}({\rm trajectory})},
\end{equation}
where $s$ is a field conjugated to the activity $C_{\tau}$. This is reminiscent of the equilibrium partition function $Z=\sum_{\rm configurations} {\rm Prob(configuration) }e^{-E/T}$ where $T$ is the temperature measured in units of the Boltzmann constant. This construction allows us to build a canonical ensemble of trajectories, conventionally named \textit{s-ensemble}. Fluctuations from the average activity correspond to particular values of the field $s$:
\begin{equation}
\langle C_{\tau}\rangle_s=\sum_{\rm trajectories} {\rm Prob(trajectory)}e^{-s C_{\tau}}C_{\tau}/Z_{C_{\tau}}
\end{equation}
with the typical average value $\bar{C}_{\tau}\equiv\sum_{\rm trajectories} {\rm Prob(trajectory)}C_{\tau}/Z_{C_{\tau}}=\langle C_{\tau}\rangle_{s=0}$ corresponding to $s=0$. We remark, however, that in the case of trajectory sampling the thermostatting temperature $T$ is not varied (it is typically chosen not too low in order to enhance sampling) because the focus is on fluctuations of time-integrated observables \textit{a-priori} unrelated to the temperature.

\begin{figure}[!htb]
\centering \includegraphics[scale=1.2]{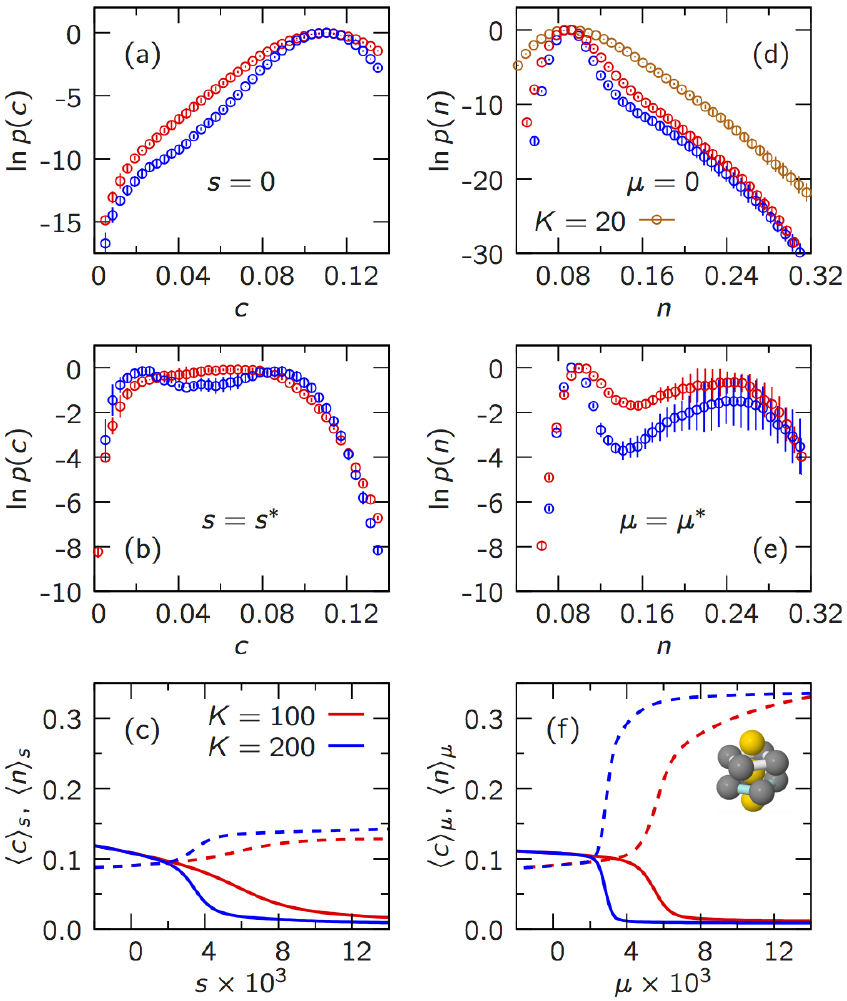} 
\caption{Phase transitions in trajectory space for the Kob-Andersen binary Lennard-Jones mixture (reproduced from~\cite{speck2012}).
Left column: $s$-ensemble (a)~Probability distributions $p(c)$ for the density of mobile particles $c$ for two trajectory lengths. The non-concave shape indicates a phase transition in trajectory space as becomes obvious from the bimodal distribution~(b) at the field $s^\ast$ that maximises the fluctuations $\langle{c^2}_s\rangle-\langle{c}_s\rangle^2$. (c)~Average fractions of mobile particles (solid lines) and bicapped square antiprism cluster population (dashed lines) \textit{vs.} the biasing field $s$. Right column: (d-f)~as left column but for the $\mu$-ensemble. Here the bicapped sqaure anitprism (depicted in (f)) is the locally favoured structure. Throughout, red and blue lines refer to $K=100$ and $K=200$, respectively. $K$ denotes the length of the trajectory. Here $K \approx 0.2 \tau_\alpha$ \cite{speck2012}.
\label{figSMu} }
\end{figure}

This framework allowed Chandler and Garrahan and co-workers \cite{hedges2009} to identify a phase transition at a coexistence value $s^{\ast}\neq0$ between an \textit{active} phase with high mobility and an \textit{inactive} phase with subextensive mobility. They used small systems ($N\sim100$ particles) of the Kob-Andersen binary mixture to demonstrate a first-order \emph{dynamical} phase transition in trajectory space.
In this \emph{dynamical} ensemble, rather than density (liquid or vapour), it is the fraction of suitably defined active (mobile) particles per trajectory $c=C_{\tau}/N$ of the trajectories which identifies the two ``phases''. The distribution in Fig. \ref{figSMu}(a) shows that low-mobility trajectories are much more likely than expected: this is taken as evidence for an ``inactive phase''. The implication is that dynamical heterogeneity is a manifestation of these two phases, fluctuating in and out of existence like density fluctuations in liquid-vapour critical phenomena.

\paragraph{The $\mathbf{\mu}$-ensemble} A formally analogous construction can be achieved if, instead of the particle activity $A_{\tau}$ on trajectories of length $\tau$, one takes the time-integrated number of particles in locally favoured structures $\mathcal{N}_{\tau}=\sum_{t=0}^{\tau} N n_\mathrm{LFS}(t)$ where $n_\mathrm{LFS}(t)$ is the concentration of locally favoured structures (LFS) in a configuration at time $t$. Locally favoured structures are recurrent local motifs whose concentration appears to increase as a glass-former is cooled down and which have been identified with emergence of slow dynamics for some systems such as hard spheres and discs or Lennard-Jones binary mixtures and metallic glassformers \cite{royall2015physrep}. The canonical ensemble of trajectories with average number of locally favoured structures per trajectory $\bar{\mathcal{N}}$ is associated to a conjugated field termed $\mu$. Again, $\mu$ is related to fluctuations from the typical value $\bar{\mathcal{N}}=\langle \mathcal{N}\rangle_{\mu=0}$  obtained at $\mu=0$, so that nonzero values of $\mu$ quantify how atypical it is to observe trajectories with time-integrated structural observable $\mathcal{N}(\mu)$.

\begin{figure}
	\centering
	\includegraphics{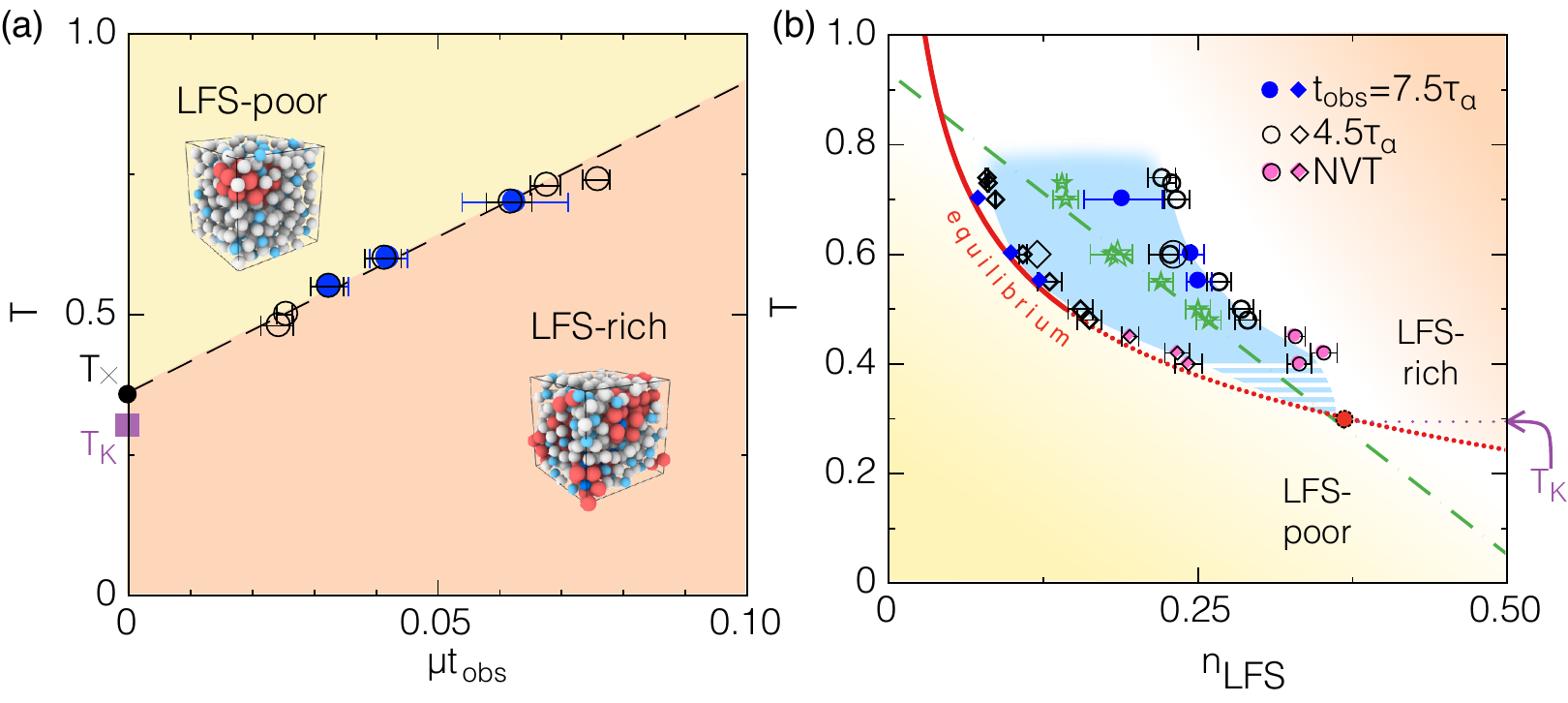}
	\caption{Results of trajectory sampling ($\mu$-ensemble) computer simulations of the Kob-Andersen binary Lennard-Jones mixture. (a) Temperature versus $\mu$ phase diagram. Two distinct structural-dynamical phases are found at coexistence at a finite value $\mu^\ast$ of the field $\mu$ when sampling trajectories of different durations $t_{\rm obs}$ (filled and empty circles): these are poor and rich in structure (LFS poor/rich), represented in the insets (with red and dark blue particles indicating the LFS regions). The scaled value $\mu^{\ast}t_{\rm obs}$, however, lies on a single curve. A linear extrapolation (dashed line) indicates that at a temperature $T_{\rm \times}\approx T_{\rm K}$ one would observe the transition from one phase to the other directly in the thermal average of structural quantities ($\mu=0$) without recurring to large deviations, under the form of intermittency. (b) Dynamical coexistence in the temperature versus concentration of LFS per trajectory plane. The coexistence region (determined by several numerical methods, in blue) has a non-trivial temperature dependence and narrows as the temperature is reduced. The equilibrium supercooled liquid approaches the coexistence region gradually and is always located close to the LFS-poor boundary. The extrapolation of the line of susceptibility maxima (green stars) and the equilibrium line meet at a temperature close to $T_{\rm K}$, suggesting a cross-over between the LFS-poor and the LFS-liquid. More information in \cite{turci2017prx}.}
	\label{figMuKA}
\end{figure}

Speck \emph{et al.} demonstrated~\cite{speck2012} that the transition in the $s$-ensemble (related to a purely dynamical observables such as the particle mobility) has its counterpart in the structural description realised in the $\mu$-ensemble: inactive trajectories correlate strongly with trajectories rich in locally favoured structures (see Fig.~\ref{figSMu}). Experimental evidence, using colloidal suspensions, has now been found for this dynamical phase transition (see Section \ref{sectionColloids} and Fig. \ref{figRattachai}) \cite{pinchiapat2017}. More recent studies~\cite{coslovich2016jstat} suggest that the inactive/locally favoured structure-rich phase obtained in the space of trajectories also correlates strongly with particularly low (potential) energy states, and that when decreasing the temperature, the inactive, LFS-rich phases tend to dominate the statistics, with the coexistence values of $s^{\ast}$ and $\mu^{\ast}$ approaching zero as the temperature is decreased. In this sense, guiding trajectory sampling with the usage of time-integrated observable can be an efficient way to identify low energy states, more present in the arrested glassy phases.

These ideas have been recently tested in the case of a canonical atomistic glass former (the Kob-Andersen binary mixture) \cite{turci2017prx}, and it has been shown that other systems exhibit similar behaviour, such as hard spheres \cite{pinchiapat2017} and the Wanhstr\"{o}m binary Lennard-Jones model \cite{turci2018}. In particular, it has been shown that the large deviations of time-integrated structural observables give access to low energy configurations that sample the tails of the probability distribution of inherent state energies. Through reweighting, this sampling recovers the thermodynamical properties of the system (such as the configurational entropy) down to very low temperatures, without the need of sampling the dynamics at low temperatures directly.

Additionally, this approach shows that the dynamical phase transition between trajectories poor/rich in structure sampled in the trajectory ensemble corresponds to a transition between two distinct metastable amorphous states at high/low inherent state energies respectively: one corresponds to the supercooled liquid sampled in conventional dynamics; the other to a secondary amorphous state, with low energy, low configurational entropy, rich in structure and very slow dynamics, see Fig.~\ref{figMuKA}. This second amorphous state is more metastable than the conventional supercooled liquid: however, the difference in stability (as measured by the value $\mu^\ast$ of the conjugated field $\mu$ at coexistence between the two phases) is a function of the temperature and \textit{decreases} as the temperature is reduced. Extrapolations indicate that $\mu^\ast(T)\rightarrow0$ at a finite temperature $T_{\times}\approx T_{\rm K}$.

This is suggestive, as it indicates the possibility of an alternative interpretation of the ideal glass transition temperature $T_{\rm K}$: at such low temperatures the structure of the liquid would change (in a continuous, or weakly discontinuous fashion) from a liquid poor in local structure to a liquid rich in structure. Such liquid-to-liquid transition would be an alternative solution of Kauzmann's original paradox, as the structure-rich liquid (whose configurational entropy is weakly dependent on the temperature, according to the simulations) would behave as a strong liquid with divergence of the relaxation times only at temperature $T=0$. We should note that these conclusions are somewhat speculative. The work suggests that the dynamical phase transition has a lower critical point close to the Kauzmann temperature. But proximity alone doesn't necessarily mean the two are connected. We discuss other scenarios for the fate of the supercooled liquid in the next section.

\section{What does it all mean?}
\label{sectionWhatDoesItAllMean}

Finally we consider what we have learnt. We began this article by noting that dynamical arrest is a major challenge and there is no firmly agreed paradigm for the phenomenon. The issue lies in understanding the behaviour of glassforming systems in the supercooling regime between the molecular glass transition $T_g$ and the Kauzmann temperature $T_K$ (or $T_0$). One might imagine that the kind of studies outlined here, namely attempts to equilibrate glassformers in this very deep range of supercooling might shed light on the nature of the glass transition. The first observation we might make is that it appears possible to supercool a system through $T_g$. In terms of experiments on molecules, which have the potential to reach states very low in the energy landscape, there is evidence that the configurational entropy continues to drop. This is  --- or \emph{should be} --- found in the work on ultrastable glasses.

In the case of numerical work, pinning (section \ref{sectionPinning}),  particle swaps (section \ref{sectionNonLocal}) and dynamical phase transitions (section \ref{sectionDynamicalPhaseTransitions}) have certainly found considerable evidence in support of a drop in configurational entropy 
\cite{ozawa2015}. This would tend to support thermodynamic theories such as Adam-Gibbs/Random First Order Transition theory. Other examples include the large deviations approach of Turci \emph{et al.} \cite{turci2017prx} which indicates a lower critical point of the dynamical transition related to dynamic facilitation theory around the Kauzmann temperature. The coming together of the equilibrium liquid and low-entropy non-equilibrium phase related to the dynamical transition suggests a possible means of unification of the large deviations approach with thermodynamic theories of Adam-Gibbs/Random First Order Transition theory \cite{adam1965,lubchenko2007} and perhaps even geometric frustration 
with a transition to a local structure-rich state albeit dynamical, rather than the avoided more conventional  transition \cite{turci2017prx}.

What can we deduce about the nature -- or existence -- or otherwise of the ideal glass, or Kauzmann transition of the Adam-Gibbs/RFOT approach? While of course this remains speculative, it is clear that, should such a transition exist, the Kauzmann point has been approached much closer than before. However, there is one piece of information lacking from much of the work which has been carried out --- and for good reason: the dynamics of such deeply supercooled materials are incredibly hard to measure. The result of not knowing the relaxation time often means we must extrapolate from higher temperature, assuming a constant fragility, with little or no justification. This means that dynamical data obtained indirectly are important. Notably two means to extract dynamical information in this regime we have mentioned above both indicate an \emph{Arrhenius} behaviour in the regime of interest, $T<T_g$ sections \ref{sectionWaitingAmber} \cite{zhao2013} and \ref{sectionUltrastable} \cite{pogna2015}.

\begin{figure}
\centering
\includegraphics{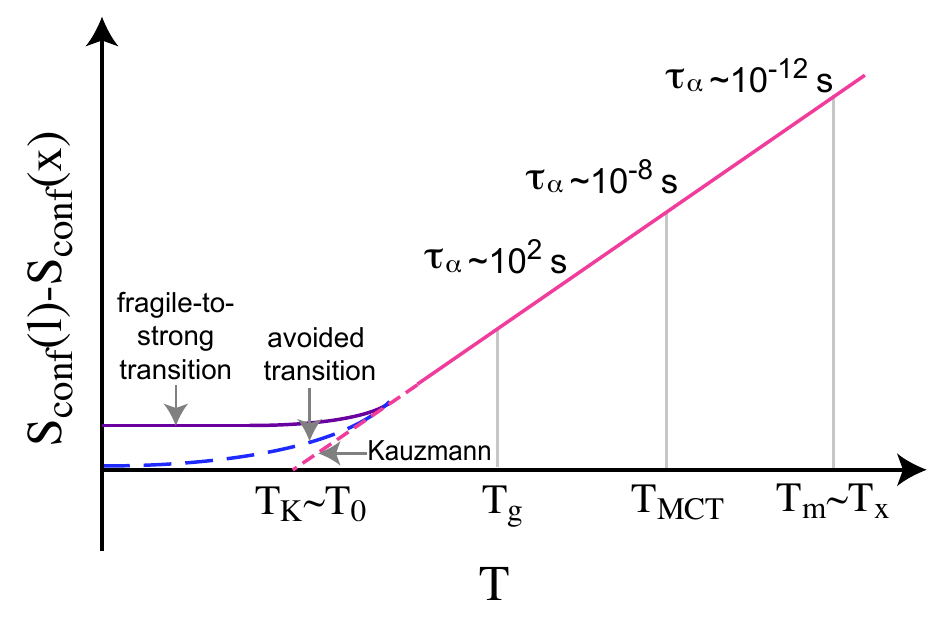}
\caption{
The decrease of the liquid configurational entropy with respect to that of the crystal. Here $S_\mathrm{conf}(l)$ and $S_\mathrm{conf}(r)$ are the liquid and crystal configurational entropies respectively. Three scenarios are depicted. The short dashed line is the original Kauzmann extrapolation~\cite{kauzmann1948}, the long dashed line pertains to the suggestion by Stillinger \emph{et al.} that any ideal glass would have ``defects'' and thus the transition would be smoothed or avoided ~\cite{stillinger2001} and the solid line is a possible fragile-to-strong transition.}
\label{figStillinger}
\end{figure}

Now such fragile-to-strong transitions have been noted before \cite{ito1999,zhang2010,royall2015physrep}, and may be related to the system reaching the bottom of its energy landscape yet somehow stopping short of the ``ideal'' glass, so that the Kauzmann transition is not reached \cite{ito1999}. 
In this picture the configurational entropy would then be essentially constant with respect to temperature, \emph{i.e.} the structure of the system is independent of temperature. This would suggest an extensive configurational entropy for all non-zero temperatures, as indicated in Fig. \ref{figStillinger}.

Such arguments relate to work by Stillinger \emph{et al.}~\cite{stillinger2001}, which are indicated in Fig. \ref{figStillinger}. Stillinger \emph{et al.} noted that that thermal motion would lead to a finite population of ``defects'' in the otherwise ``ideal'' glass, thus making the transition to any ideal glass into a crossover at best. This would lead to a finite relaxation time. Indeed, if this ``defect population" were somehow constant over some temperature range, even Arrhenius behaviour might be found.

\emph{Perspectives. --- } After the above arguments, it may seem that we are little closer to resolving the origin of the glass transition. However, while the origin of the slow dynamics remains out of reach for now, it feels tantalisingly close. We close with the following observations.
\begin{itemize}
  \item Adam-Gibbs/RFOT-like behaviour is found until very close to the point at which the Kauzmann transition is predicted.
  \item It seems possible to unify the seemingly different approaches of dynamic facilitation and Adam-Giibs/RFOT via the discovery of a possible lower critical point to the dynamical phase transition close to or coincident with the Kauzmann temperature.
  \item If any insight can be gained as to the dynamics in the new, deeply supercooled simulation data, it would be most useful. The methods of Pogna \emph{et al.} \cite{pogna2015} might be taken as a starting point, though the challenges of obtaining suitable data are hard to overstate.
  \item Some means to identify ``defects'' such as those postulated by Stillinger \emph{et al.}~\cite{stillinger2001} might provide another route to deducing which scenario in Fig.  \ref{figStillinger} holds. Some attempt to identify defects in the form of ``locally unfavoured structures'' has been made \cite{royall2017,dunleavy2015}, but much remains to be done. Order agnostic approaches such as overlap may provide crucial insight here~\cite{kob2013}. 
   \item If we can somehow deduce the dynamics, then the more ambitious task of identifying dynamic lengthscales beyond $T_g$ may well provide key evidence to discriminate between competing theoretical approaches.
\end{itemize}

Thus we argue that the remarkable progress made in the last few years towards the bottom of the energy landscape sets us up to refine the approach we might take towards resolving the existence -- or otherwise -- of the ideal glass. In particular we believe that the key lies in probing dynamics, and if possible, dynamic lengthscales, and ``defects''.

\vspace{2cm}
\ack
In connection with the preparation of this article, the authors would like to thank 
Chistiane Alba-Simionesco,
Austen Angell, 
Ludovic Berthier, 
Giulio Biroli,
Chiara Cammarota,
David Chandler, 
Patrick Charbonneau,
Eric Corwin,
Daniele Coslovich,
Olivier Dauchot,
Jeppe Dyre,
Mark Ediger, 
Juan P. Garrahan,
Peter Harrowell,
Trond Ingebrigsten, 
Rob P. Jack,
Walter Kob,
Andrea Liu,
Tannie Liverpool,
Kuni Miyazaki,
Mike Moore,
Sid Nagel,
Ross Poldark,
Itamar Procaccia, 
Tulio Sciopignio,
Francesco Sciortino,
Thomas Speck,
Grzegorz Szamel,
Hajime Tanaka,
Gilles Tarjus,
David Wales,
Eric Weeks,
Stephen Williams,
Mathieu Wyart,
Hai-Bin Yu,
Emanuela Zaccarelli,
and 
Francesco Zamponi
for helpful discussions. We thank Ludovic Berthier, Daniele Coslovich, Patrick Charbonneau and Mark Ediger for helpful comments on the manuscript.
Rattachai Pinchaipat and Sho Yaida are gratefully acknowledged for providing data.
Chiharu Nakamura is thanked for the bally-ball stuff.
CPR gratefully acknowledges the Royal Society, European Research Council (ERC Consolidator Grant NANOPRS, project number 617266) and Kyoto University SPIRITS fund for financial support.  


\end{document}